\documentclass[12pt]{article}
\usepackage{epsfig}
\usepackage{graphics}
\newcommand{\comment}[1]{}
\def\be{\begin{equation}}
\def\ee{\end{equation}}
\def\beq{\begin{equation}}
\def\eeq{\end{equation}}
\def\bea{\begin{eqnarray}}
\def\eea{\end{eqnarray}}

\def\gv{g_V}
\def\ga{g_A}
\def\rev1{{\rm Re}\ V_1}
\def\rea1{{\rm Re}\ A_1}
\def\imv1{{\rm Im}\ V_1}
\def\ima1{{\rm Im}\ A_1}
\def\rev2{{\rm Re}\ V_2}
\def\rea2{{\rm Re}\ A_2}
\def\imv2{{\rm Im}\ V_2}
\def\ima2{{\rm Im}\ A_2}
\def\rev3{{\rm Re}\ V_3}
\def\rea3{{\rm Re}\ A_3}
\def\imv3{{\rm Im}\ V_3}
\def\ima3{{\rm Im}\ A_3}

\def\eehz{$e^+e^-\to ZH$~}

\begin{document}

\begin{center}
\boldmath
{\Large \bf Charged lepton distributions as a probe of \\[2mm] 
contact $e^+e^-HZ$
interactions \\[4mm] at a linear collider with polarized beams
}
\vskip 1cm
{\large Kumar Rao and Saurabh D. Rindani}\\
\smallskip
\smallskip
{\it Theoretical Physics Division, Physical Research Laboratory \\
Navrangpura, Ahmedabad 380009, India}
\vskip 2cm
{\bf Abstract}
\end{center}

\begin{quote}
We examine very general four-point interactions arising from new
physics and contributing to the Higgs production 
process $e^+e^- \to HZ$. We write all 
possible forms for these interactions consistent with Lorentz
invariance. We allow the possibility of CP violation. 
Contributions to the process from anomalous $ZZH$ and
$\gamma ZH$ interactions studied earlier arise as a special
case of our four-point amplitude. We consider the decay of $Z$ into a
charged lepton pair and obtain expressions for 
angular distributions of charged leptons arising from the interference
of the four-point contribution with the standard-model contribution. 
We take into account possible longitudinal or transverse beam
polarization likely to be available at a linear collider. 
We examine several correlations which can be used to study the various
form factors present in the $e^+e^-HZ$ contact interactions.
We also obtain the sensitivity of these correlations 
in constraining the new-physics interactions at a linear collider
operating at a centre-of-mass energy of 500 GeV with longitudinal 
or transverse polarization.
\end{quote}

\vskip 2cm

\section{Introduction}

Despite the dramatic success of the standard model (SM), 
an essential component 
of SM responsible for generating masses in the theory, viz., the Higgs 
mechanism, as yet remains untested. The SM Higgs boson, signalling
symmetry breaking in SM by means of one scalar doublet of $SU(2)$, is
yet to be discovered. A scalar boson with the properties of the SM Higgs
boson is likely to be discovered at the Large Hadron Collider
(LHC). However, there are a number of scenarios beyond the standard
model for spontaneous symmetry
breaking, and ascertaining the mass and other properties of the
scalar boson or bosons is an important task. This task would prove
extremely difficult for LHC. However, scenarios beyond SM, with 
more than just one Higgs doublet, as in the case of minimal supersymmetric
standard model (MSSM), would be more amenable to discovery at a linear $e^+e^-$
collider operating at a centre-of-mass (cm) energy of 500 GeV. We are at
a stage when the International
Linear Collider (ILC), seems poised to become a reality \cite{LC_SOU}. 

Scenarios going beyond the SM mechanism of symmetry breaking, and
incorporating new mechanisms of CP violation, have also become a
necessity in order to understand baryogenesis which resulted in the
present-day baryon-antibaryon asymmetry in the universe. In a theory
with an extended Higgs sector and new mechanisms of CP violation, the
physical Higgs bosons are not necessarily eigenstates of CP. 
In such a case, the production of a physical Higgs can proceed through
more than one channel, and the interference between two channels can
give rise to a CP-violating signal in the production.

\begin{figure}[t]
\begin{center}
\epsfig{height=5cm,file=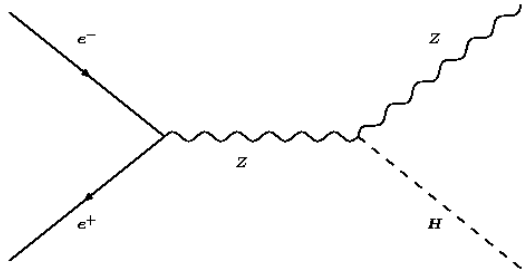}
\end{center}
\caption{Higgs production diagram with an $s$-channel exchange of $Z$
with point-like $ZZH$ coupling.}
\label{fig:vvhptgraph}
\end{figure}

\begin{figure}[t]
\begin{center}
\epsfig{height=4cm,file=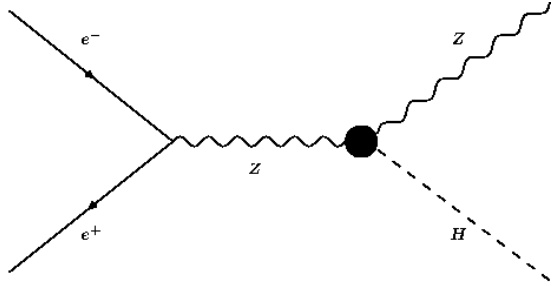}
\end{center}
\caption{Higgs production diagram with an $s$-channel exchange of $Z$
with anomalous $ZZH$ coupling.}
\label{fig:vvhgraph}
\end{figure}

\begin{figure}[t]
\begin{center}
\epsfig{height=4cm,file=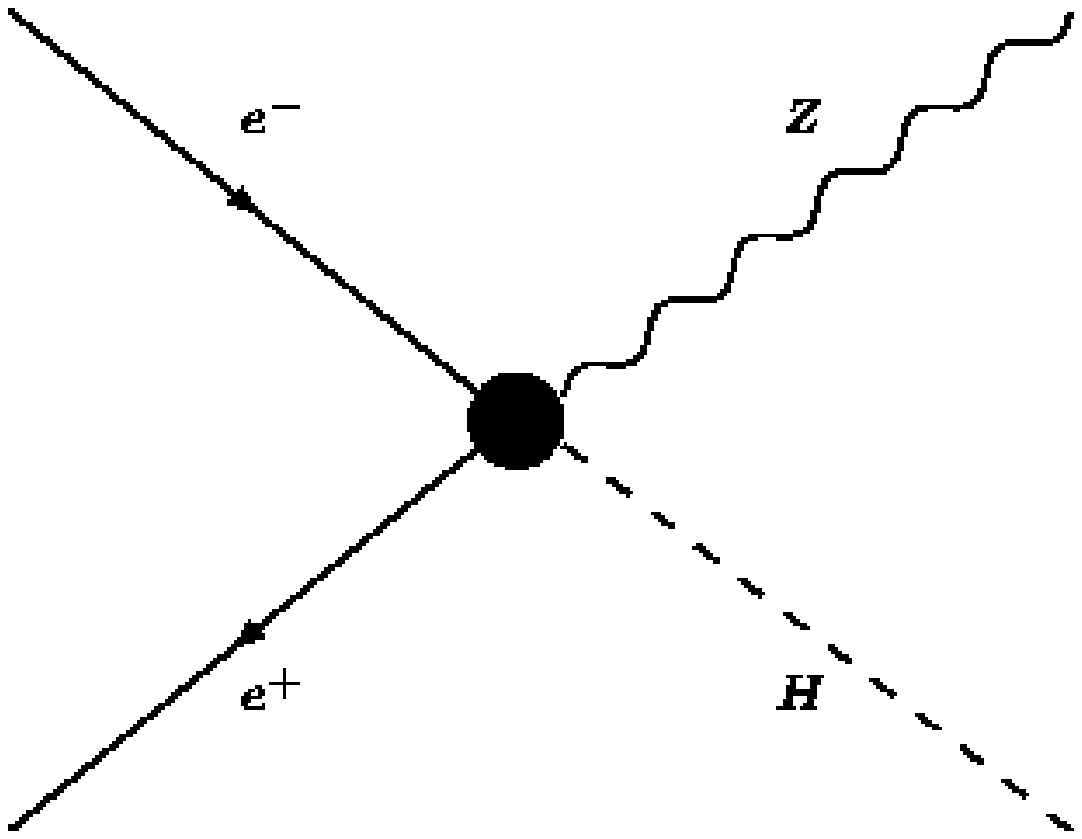}
\end{center}
\caption{Higgs production diagram with a four-point coupling.} 
\label{fig:eehzgraph}
\end{figure}

Here we consider in a general model-independent way the production of a
Higgs mass eigenstate $H$ through the process $e^+e^- \to HZ$. This is
an important mechanism for the production of the Higgs, the other
important mechanisms being $e^+e^- \to e^+e^- H$ and $e^+e^- \to \nu
\overline \nu H$ proceeding via vector-boson fusion. $e^+e^- \to HZ$ is
generally assumed to get a contribution from a diagram with an 
$s$-channel exchange of $Z$. At the lowest
order, the $ZZH$ vertex in this diagram would be simply a point-like
coupling (Fig. \ref{fig:vvhptgraph}).  Interactions
beyond SM can modify this point-like vertex by means of a
momentum-dependent form factor,
as well as by adding more complicated momentum-dependent 
forms of anomalous interactions considered
in \cite{cao,biswal,han,zerwas,gounaris,hagiwara,kniehl1,kniehl2}. 
The corresponding diagram
is shown in Fig. \ref{fig:vvhgraph}, where the anomalous $ZZH$ vertex is
denoted by a blob. There could also be a diagram with a photon
propagator and an anomalous $\gamma ZH$ vertex, which does not occur
in SM at tree level, and which we do not show separately.
We consider here a beyond-SM
contribution represented by a four-point coupling shown in Fig.
\ref{fig:eehzgraph}. This is general enough to include the effects of
the diagram in Fig. \ref{fig:vvhgraph}. Such a discussion would be
relevant in studying effects of box diagrams with new particles, 
or diagrams with $t$-channel exchange of new particles, in addition to
$s$-channel diagrams.

We write down the most general form for the four-point 
coupling consistent with Lorentz invariance. We do not assume CP
conservation.
We then obtain angular distributions for a lepton pair from the 
decay of the $Z$  
calculated from the square of amplitude $M_1$ for the diagram in Fig.
\ref{fig:vvhptgraph} with a point-like $ZZH$ coupling, 
together with  the cross term between $M_1$ and the
amplitude $M_2$ for the diagram in Fig. \ref{fig:eehzgraph}. We neglect the square of
$M_2$, assuming that this new physics contribution is small compared to
the dominant contribution $\vert M_1 \vert^2$.
We include the possibility that the beams have polarization, either
longitudinal or transverse.
While we have restricted the actual calculation to SM couplings in
calculating $M_1$, it should be borne in mind that 
in models with more than one Higgs doublet this
amplitude would differ by an overall factor depending on the mixing
among the Higgs doublets. Thus our results are trivially applicable to
such extensions of SM, by an appropriate rescaling of the coupling.

In our analysis, we do not assume that the $Z$ is produced on shell.
Moreover, since we obtain fully analytical expressions for the distribution
of the final-state leptons arising from the virtual $Z$ using the full
$Z$ propagator, we
automatically take into account coherently spin correlations of the $Z$.

We are thus addressing the question of how well the form factors for the
four-point $e^+e^-HZ$ coupling can be determined from the observation of
decay-lepton angular distributions in the presence of unpolarized beams or beams
with either longitudinal or transverse polarizations. A similar question
taking into account a new-physics contribution which merely modifies the
form of the 
$ZZH$ vertex has been addressed before in several works
\cite{cao,biswal,han,zerwas,gounaris,hagiwara,kniehl1,kniehl2,Skjold}. Those 
works which do take into account four-point couplings, do not do so in all 
generality, but
stop at the lowest-dimension operators \cite{zerwas}. See, however,
\cite{kile}, where the contribution of dimension-six operators to the
processes $e^+e^- \to H l^+l^-$ and $e^+e^- \to H \nu \bar \nu$ are
considered.  
The approach we adopt here has been used for the process $e^+e^- \to
\gamma Z$ in \cite{basdr,AL1} and for the process $Z\to b \overline b
\gamma$ in \cite{AL2}.

The four-point couplings, in the limit of vanishing electron mass, can
be neatly divided into two types -- chirality-conserving (CC) ones and
chirality-violating (CV) ones. The CC couplings involve an odd number of
Dirac $\gamma$ matrices sandwiched between the electron and positron spinors,
whereas the CV ones come from an even number of Dirac $\gamma$ matrices.
Since in practice, CV couplings are usually
proportional to the fermionic mass (in this case the electron mass),  
we concentrate on the CC ones (see, however, \cite{lepto}). 

In an earlier work \cite{plb}, we considered angular distributions of an 
on-shell $Z$ in the process \eehz in the same context of $e^+e^-HZ$ 
contact interactions. In that paper we concentrated on CP-odd
asymmetries constructed with the $Z$ polar and azimuthal angles, for
both chirality-conserving and chirality-violating cases. The
present work is an extension to the more realistic case of a virtual
$Z$ decaying into a pair of charged leptons, which are observed. We also
do not restrict ourselves to CP-odd asymmetries, but aim at the
determination of all form factors in the chirality-conserving case using
expectation values of CP-even and CP-odd observables. 
Charged-lepton angular distributions have been discussed earlier in
\cite{chen} for the SM Higgs, in the context of 
distinguishing between a scalar and pseudoscalar Higgs in 
\cite{cao,hagiwara,kniehl1,Skjold,mahlon}, and in the context of CP-violating Higgs
in \cite{biswal,han,zerwas,gounaris,hagiwara,kniehl2,Skjold}. 
However, these papers, with the exception of \cite{zerwas}, do not discuss contact
interactions which is the topic of our work.

Polarized beams are likely to be available at a linear collider, and
several studies have shown the importance of linear
polarization in reducing backgrounds and improving the sensitivity to
new effects \cite{gudi}. The question of whether transverse beam
polarization, which could be obtained with the use of spin rotators,
would be useful in probing new physics, has been addressed in recent
times in the context of the ILC 
\cite{basdr,lepto,gudi,rizzo,basdrtt,basdrzzg,bartl,Rindani:2004wr}.
In earlier work, it has been observed that polarization does not give
any new information about the anomalous $ZZH$ couplings when they are
assumed real \cite{hagiwara}. 
However, in case of four-point contact interactions, 
we find that there are terms in the differential cross section
which are absent unless both electron and positron beams are transversely
polarized. Thus, transverse polarization, if available at ILC, would be
most useful in isolating such terms. This is particularly significant
because these terms are CP violating. Moreover, some of them are 
even under naive
CPT, and thus would survive even when no imaginary part is present in
the amplitude. 

In the next section we write down the possible model-independent
four-point $e^+e^-HZ$ couplings. In Section 3, we obtain the angular
distributions arising from the CC couplings in the presence of beam
polarization. Section 4 deals with correlations which can be used
for separating various form factors and Section 5 describes the numerical
results.  Section 6
contains our conclusions and a discussion.

\section{\boldmath Form factors for the process $e^+e^- \to HZ$}

The most general four-point vertex for the process 
\begin{equation}\label{process}
e^- (p_1) + e^+ (p_2) \to Z^\alpha (q) + H(k)
\end{equation}
consistent with Lorentz invariance can be written as
\begin{equation}\label{vertices}
\Gamma^\alpha_{\rm 4pt} =  \Gamma^\alpha_{\rm CC} + 
				\Gamma^\alpha_{\rm CV},
\end{equation}
where the chirality-conserving part $\Gamma^\alpha_{\rm CC}$
containing an odd number of Dirac $\gamma$ matrices is
\def \slZ{Z{\hskip -1.5ex}/}
\def \slq{q{\hskip -1.2ex}/}
\def \slpart{\partial{\hskip -1.25ex}/}
\def \dsp{\displaystyle}

\comment{ This is the old equation:
\begin{equation}\label{vertexcc}
\Gamma^\alpha_{\rm CC} = \frac{i}{M} \gamma^\alpha (V_1 + \gamma_5 A_1)
			-\frac{i}{M^3}\slq (V_2 + \gamma_5 A_2) k^\alpha
			-\frac{1}{M^3}\slq (V_3 + \gamma_5 A_3) (p_2 - p_1)^ 
				\alpha ,
\end{equation}
}
\begin{equation}\label{vertexcc}
\Gamma^\alpha_{\rm CC} = -\frac{1}{M} \gamma^\alpha (V_1 + \gamma_5 A_1)
			+\frac{1}{M^3}\slq (V_2 + \gamma_5 A_2) k^\alpha
			-\frac{i}{M^3}\slq (V_3 + \gamma_5 A_3) (p_2 - p_1)^ 
				\alpha ,
\end{equation}
and the chirality violating part containing an even number of Dirac
$\gamma$ matrices is
%\begin{equation}
\comment{This is the old equation
\begin{eqnarray}\label{vertexcv}
\Gamma^\alpha_{\rm CV}& =&\dsp \frac{1}{M^2}\left[- (S_1 + i \gamma_5 P_1)
				k^\alpha + i (S_2 + i \gamma_5 P_2) 
				(p_2 - p_1)^\alpha \right]\nonumber\\
			&&\!\!\!\! +\dsp
				\frac{i}{M^4}
			\epsilon^{\mu\nu\alpha\beta}p_{2\mu}p_{1\nu}k_\beta 
			(S_3 + i \gamma_5 P_3) .
\end{eqnarray}
}
\begin{eqnarray}\label{vertexcv}
\Gamma^\alpha_{\rm CV}& =&\dsp \frac{i}{M^2}\left[- (S_1 + i \gamma_5 P_1)
				k^\alpha -  (S_2 + i \gamma_5 P_2) 
				(p_2 - p_1)^\alpha \right]\nonumber\\
			&&\!\!\!\! -\dsp
				\frac{1}{M^4}
			\epsilon^{\mu\nu\alpha\beta}p_{2\mu}p_{1\nu}k_\beta 
			(S_3 + i \gamma_5 P_3) .
\end{eqnarray}
%\end{equation}
In the above expressions, $V_i$, $A_i$, $S_i$ and $P_i$ are form
factors, and are Lorentz-scalar functions of the Mandelstam variables
$s$ and $t$ for the process eq. (\ref{process}). For simplicity, we will only 
consider the case here when the form factors are constants. $M$ is a
parameter with dimensions of mass, put in to render the form factors
dimensionless.

The expressions for the four-point vertices may be thought to arise from
effective Lagrangians
%\begin{equation}
\begin{eqnarray}\label{Lcc}
{\cal L}_{\rm CC}& =& 
			\frac{1}{M} \bar \psi \slZ (v_1 + \gamma_5 a_1) 
                        \psi \phi \nonumber\\
	&&\!\!\!\!		+
			\frac{1}{M^3} \bar \psi \slpart Z^\alpha 
			(v_2 + \gamma_5 a_2) 
                        \psi \partial_\alpha \phi \nonumber \\
	&&\!\!\!\!	+  \frac{i}{M^3} \left[\partial_\alpha \bar \psi
                        \gamma^\mu (v_3 + \gamma_5 a_3) \psi 
			- \bar \psi \gamma^\mu (v_3 + \gamma_5 a_3)
			  \partial_\alpha\psi \right]
                         \phi \partial_\mu Z^\alpha ,
\end{eqnarray}
%\end{equation}
and
%\begin{equation}
\begin{eqnarray}\label{Lcv}
{\cal L}_{\rm CV}& =&  \frac{1}{M^2} \bar \psi  (s_1 + i\gamma_5 p_1) 
                        \psi \partial_\alpha\phi Z^\alpha\nonumber\\
	&&\!\!	\!\!	+  \frac{i}{M^2} \left[\partial_\alpha \bar \psi
                         (s_2 +i \gamma_5 p_2) \psi 
			- \bar \psi  (s_2 + i \gamma_5 p_2)
			  \partial_\alpha\psi \right] \phi
				Z^\alpha \nonumber\\
 	&&\!\!\!\!		+ 
			\frac{i}{M^4} \epsilon^{\mu\nu\alpha\beta} \partial_\mu 
			\bar \psi (s_3 + i \gamma_5 p_3) \partial_\nu \psi 
			\partial_\beta \phi Z_\alpha  ,
\end{eqnarray}
%\end{equation}
where $\phi$ represents the Higgs field. The 
the coupling constants $v_i$, $a_i$, $s_i$ and $p_i$ in the
Lagrangians have been promoted to form factors in momentum space when
writing the vertex functions $\Gamma$.

It may be appropriate to contrast our approach with the usual effective
Lagrangian approach. In the latter approach, it is
assumed that SM is an effective theory which is valid up to a cut-off scale
$\Lambda$. The new physics occurring above the scale of the cut-off may
be parametrized by higher-dimensional operators, appearing with powers
of $\Lambda$ in the denominator. These when added to the SM Lagrangian
give an effective low-energy Lagrangian
where, depending on the scale of the momenta involved, one
includes a range of higher-dimensional operators up to a certain maximum
dimension. Our effective theory is not a low-energy limit, so that the
form factors we use are functions of momentum not restricted to low
powers. Thus, the $M$ we introduce is not a cut-off scale, but an
arbitrary parameter, introduced just to make the form factors
dimensionless. 

We thus find that there are 6 independent form factors in the chirality
conserving case, and 6 in the chirality
violating case. An alternative form for the $\Gamma$ above would be
using Levi-Civita $\epsilon$ tensors whenever a $\gamma_5$ occurs. The
independent form factors then are then some linear combinations of the
form factors given above. However, the total number of independent form
factors remains the same.

Note that we have not imposed CP conservation in the above. 
The CP properties of the various terms appearing in the four-point
vertices may be deduced from the CP properties of the corresponding
terms in the effective Lagrangian. Thus, one can check that the terms
corresponding to the couplings $v_3$, $a_3$, $s_1$, $p_2$ and $s_3$ in
the effective Lagrangian are CP violating. As a consequence, the 
terms corresponding to 
$V_3$, $A_3$, $S_1$, $P_2$ and $S_3$ are CP violating, whereas the
remaining are CP conserving. This conclusion assumes that the form factors are 
constants, since the couplings in the effective Lagrangian are
constants. The conclusion can also be carried over when the form factors
are arbitrary functions of $s$ and even
functions of $t-u\equiv \sqrt{s}\vert \vec q \vert \cos\theta$, where
$\theta$ is the angle between $\vec q$ and $\vec p_1$ (or constants). 
This is because in momentum space, $s\equiv (p_1+p_2)^2$ is even under CP,
whereas $t-u\equiv \sqrt{s}\vert \vec q \vert \cos\theta$ is odd under
C and even under P, and thus odd under CP.

\section{Differential cross sections}

We now obtain the differential cross section for the process
(\ref{process}) for a virtual $Z$ followed by its decay into $l^+l^-$,
viz., 
\be\label{decay}
Z(q) \to l^+(p_{l^+}) + l^-(p_{l^-}),
\ee
from the SM amplitude alone and from the interference between the SM
amplitude and the amplitude arising from the four-point couplings of
(\ref{vertexcc}). We do not consider the CV case here. We ignore terms
bilinear in the four-point couplings, assuming that the new-physics
contribution is small. We treat the two cases of longitudinal and
transverse polarizations for the electron and positron beams separately.
We neglect the mass of the charged leptons $l^{\pm}$. Also, we assume
that the charged lepton $l$ is different from the electron. Thus, our
considerations are mostly for $l\equiv \mu$, and they would be
applicable for $l\equiv \tau$ to the extent that the $\tau$ mass can be
neglected.

The expression for the amplitude for (\ref{process}), arising from the SM 
diagram of Fig. \ref{fig:vvhptgraph} with a point-like $ZZH$ vertex, is 
\begin{equation}\label{smamp}
M_{\rm SM} =- \frac{e^2}{4\sin^2\theta_W \cos^2\theta_W}
\frac{m_Z}{s-m_Z^2} \overline v (p_2) \gamma^\alpha (\gv - \gamma_5 \ga)
u(p_1),
\end{equation}
where the vector and axial-vector couplings of the $Z$ to electrons are
given by
\beq\label{gvga}
\gv= -1 + 4 \sin^2\theta_W,\; \ga=-1,
\eeq
and $\theta_W$ is the weak mixing angle. 

Note that though we have used SM couplings for the leading contribution,
it is trivial to modify these by overall factors for cases of other
models (like two-Higgs-doublet models). Our expressions are not, however, 
applicable for the case when the Higgs is a pure pseudoscalar in models
conserving CP, since in that case, the SM-like lowest-order couplings
are absent. 
 
We have obtained full analytical expressions for the differential cross
sections to linear order in the contact interactions. 
The Dirac trace calculations have
been checked using the analytical manipulation program FORM \cite{form}.
Since these expressions are obtained in a model-independent context,
they can prove useful for future work comparing different models.

Note that though in eq. (\ref{vertexcc}) 
we wrote the vertex for the production of a real
$Z$, when we introduce the decay of the $Z$, 
we consider full virtuality of the $Z$, and also take into account
spin correlations of the $Z$.

We choose the $z$ axis to be the direction of the $e^-$ momentum, and
the $xz$ plane to coincide with the $HZ$ production plane in the case
when the initial beams are unpolarized or longitudinally polarized. 
The positive $x$
axis is chosen, in the case of transverse polarization, to be along the
direction of the $e^-$ polarization. 
%We then define $\theta$ and $\phi$
%to be the polar and azimuthal angles of the momentum $\vec q$ of the $Z$.
%Similarly we define $\theta_{l^{\pm}}$ and $\phi_{l^{\pm}}$
%to be the polar and azimuthal angles of the momenta $p_{l^{\pm}}$ of the
%charged leptons $l^{\pm}$ arising from the decay of the $Z$.  

\subsection{Distributions for longitudinally polarized beams}

The cross
section for the process $e^+e^- \to l^-l^+ H$ 
for longitudinal beam polarization is given by
\begin{eqnarray}\label{longcs}
\sigma_{\rm L}& =& \int \frac{d^3 p_{l^-}}{2p^0_{l^-}} \int \frac{d^3
p_{l^+}}{2p^0_{l^+}} 
\left(\frac{e}{4 \sin \theta_W \cos \theta_W}\right)^2 
\frac{1}{(q^2 -m_Z ^2)^2 +\Gamma_Z ^2 m_Z^2}
(1-P_L \overline{P}_L) 
\nonumber \\
&&\times
\left[ \mathcal{F}^{\rm L}_{\rm SM} + 
\mathcal{F}^{\rm L}_1+
\mathcal{F}^{\rm L}_2+
\mathcal{F}^{\rm L}_3 \right].
\end{eqnarray}
Here, respectively $\mathcal{F}^{\rm L}_{\rm SM}$ and  $\mathcal{F}^{\rm
L}_i$ ($i=1,2,3$)
represent contributions of the pure SM
amplitude and interference between the SM amplitude and the amplitudes
coming from the couplings $V_i$, $A_i$ ($i=1,2,3$).
The expressions for $\mathcal{F}^{\rm L}_{\rm SM}$ and  $\mathcal{F}^{\rm
L}_i$ are as
follows. We use the notation $q=p_{l^-}+p_{l^+}$, as already specified,
and $P=p_1+p_2$.
%%%%%%%%%%%%%%%%%%%%%%%
{\setlength\arraycolsep{2pt}
\begin{eqnarray}
\mathcal{F}^{\rm L}_{\rm SM} &=&8 F^2 \bigg[(g_V^2 -g_A ^2)^2 (p_1 \cdot p_{l^-}) (p_2 \cdot p_{l^+}) \nonumber \\
&& \hskip -1.8cm +  \left\{((g_V^2 + g_A^2)^2+ 4g_A^2 g_V^2) -4
P_{L}^{\rm{eff}}(g_V^2 +g_A^2)g_V g_A \right\}(p_1 \cdot p_{l^+}) (p_2
\cdot p_{l^-})\bigg],
\end{eqnarray}}
%%%%%%%%%%%%%%%%%%%%%%%
{\setlength\arraycolsep{2pt}
\begin{eqnarray}
\mathcal{F}^{\rm L}_1 &=& \frac{16F}{M}\bigg[(g_V^2 +g_A ^2)\left\{(g_V -g_A  P_{L}^{\rm{eff}})\textrm{Re}V_1+(g_V  P_{L}^{\rm{eff}}-g_A)\textrm{Re}A_1\right\}   \nonumber \\ 
& &{}\times \left\{(p_1 \cdot p_{l^-}) (p_2 \cdot p_{l^+})+(p_1 \cdot p_{l^+}) (p_2 \cdot p_{l^-})\right\} \nonumber \\
& &{}+2 g_V g_A \left\{(g_V -g_A  P_{L}^{\rm{eff}})\textrm{Re}A_1+(g_V  P_{L}^{\rm{eff}}-g_A)\textrm{Re}V_1 \right\} \nonumber \\
& & \times \left\{(p_1 \cdot p_{l^-}) (p_2 \cdot p_{l^+})-(p_1 \cdot
p_{l^+}) (p_2 \cdot p_{l^-})\right\} \bigg],
\end{eqnarray}}
%%%%%%%%%%%%%%%%%%%%%%%%%%
{\setlength\arraycolsep{2pt}
\begin{eqnarray}
\mathcal{F}^{\rm L}_2
 &=&- \frac{8F}{M^3}\bigg[(g_V^2 +g_A ^2)\left\{(g_V -g_A  P_{L}^{\rm{eff}})\textrm{Re}V_2 +(g_V  P_{L}^{\rm{eff}}-g_A)\textrm{Re}A_2 \right\}   \nonumber \\ 
&& \hskip -.4cm \times \left\{ p_1 \cdot q\left[(p_2 \cdot p_{l^-})P\cdot p_{l^+}+(p_2 \cdot p_{l^+})P\cdot p_{l^-}-(p_{l^-}\cdot p_{l+})\frac{s}{2}\right] \right. \nonumber \\
&&\hskip -.4cm + \left. p_2 \cdot q \left[(p_1 \cdot p_{l^-})P\cdot p_{l^+}+(p_1 \cdot p_{l^+})P\cdot p_{l^-}-(p_{l^-}\cdot p_{l+})\frac{s}{2}\right] \right\} \nonumber \\
& &{}\hskip -.4cm -2 g_V g_A \left\{ (g_V  -g_A  P_{L}^{\rm{eff}})\textrm{Im}V_2 +(g_V  P_{L}^{\rm{eff}}-g_A)\textrm{Im}A_2 \right\}   \nonumber \\ 
& &{}\hskip -.4cm \times (p_2 -p_1)\cdot q \,\,\epsilon_{\alpha \beta
\sigma \rho}\,p_1^{\alpha}\,p_2^{\beta}\,p_{l^-}^{\sigma}\,p_{l^+}^{\rho}  
\nonumber \\
& &{}\hskip -.4cm+ (g_V^2 +g_A^2)\left\{(g_A  -g_V  P_{L}^{\rm{eff}})\textrm{Im}V_2 +(g_A  P_{L}^{\rm{eff}}-g_V)\textrm{Im}A_2 \right\}   \nonumber \\ 
& &{}\hskip -.4cm\times  P\cdot (p_{l^+}-p_{l^-})\,\, \epsilon_{\alpha
\beta \sigma
\rho}\,p_{1}^{\alpha}\,p_{2}^{\beta}\,p_{l^-}^{\sigma}\,p_{l^+}^{\rho}  
\nonumber \\
& &{}\hskip -.4cm-2 g_V g_A \left\{ (g_V  -g_A  P_{L}^{\rm{eff}})\textrm{Re}A_2 +(g_V  P_{L}^{\rm{eff}}-g_A)\textrm{Re}V_2 \right\}   \nonumber \\ 
&&\hskip -.4cm \times \left\{ (p_{l^-}\cdot p_{l^+})\left[\left(\frac{s}{2}-p_2\cdot q\right)p_1 \cdot (p_{l^-}-p_{l^+})-\left(\frac{s}{2}-p_1\cdot q\right) p_2 \cdot (p_{l^-}-p_{l^+})\right] \right. \nonumber \\
&&\hskip -.4cm + (P\cdot q -2 p_{l^-}\cdot p_{l^+})\left[(p_1
\cdot p_{l^+})(p_2 \cdot p_{l^-})-(p_1 \cdot p_{l^-})(p_2 \cdot
p_{l^+})\right] \bigg\} \bigg],
\end{eqnarray}}
%%%%%%%%%%%%%%%%%%%%%%%
{\setlength\arraycolsep{2pt}
\begin{eqnarray}
\mathcal{F}^{\rm L}_3 
&=& \frac{8F}{M^3}\bigg[-(g_V^2 +g_A ^2)\left\{(g_V-g_A P_{L}^{\rm{eff}})\textrm{Im}V_3+(g_V P_{L}^{\rm{eff}}-g_A) \textrm{Im}A_3 \right\} \nonumber \\
&& \hskip 0cm \times \bigg\{(p_2 -p_1)\cdot p_{l^+} [(p_1 \cdot p_{l^-})(p_2 \cdot q)+ (p_1 \cdot q)(p_2 \cdot p_{l^-})] \nonumber \\
&&  \hskip 0cm +(p_2 -p_1)\cdot p_{l^-} [(p_1 \cdot p_{l^+})(p_2 \cdot q)+ (p_1 \cdot q)(p_2 \cdot p_{l^+})] \nonumber \\
&& \hskip 0cm-\frac{s}{2}(p_2 -p_1)\cdot q (p_{l^-}\cdot p_{l^+})\bigg\} \nonumber \\
&& \hskip 0cm+2 g_V g_A \left\{(g_V-g_A P_{L}^{\rm{eff}})\textrm{Im}A_3+(g_V P_{L}^{\rm{eff}}-g_A) \textrm{Im}V_3 \right\} \nonumber \\
&& \hskip 0cm \times \bigg\{(p_2 -p_1)\cdot q [(p_1 \cdot p_{l^+})(p_2 \cdot p_{l^-})- (p_1 \cdot p_{l^-})(p_2 \cdot p_{l^+})] \nonumber \\
&& \hskip 0cm+\frac{s}{2}P\cdot (p_{l^+}-p_{l^-}) (p_{l^-}\cdot p_{l^+})\bigg\} \nonumber \\ 
&& \hskip 0cm-2 g_V g_A \left\{(g_V-g_A P_{L}^{\rm{eff}})\textrm{Re}V_3+(g_V P_{L}^{\rm{eff}}-g_A) \textrm{Re}A_3 \right\} \nonumber \\
&& \hskip 0cm\times P\cdot q\,\, \epsilon^{\alpha \beta \sigma \rho}\,p_{1\alpha}\,p_{2\beta}\,p_{l^- \sigma}\,p_{l^+ \rho}  \nonumber \\
&& \hskip 0cm+(g_V^2 +g_A^2) \left\{(g_V-g_A P_{L}^{\rm{eff}})\textrm{Re}A_3+(g_V P_{L}^{\rm{eff}}-g_A) \textrm{Re}V_3 \right\} \nonumber \\
&& \hskip 0cm\times (p_2-p_1)\cdot (p_{l^-}-p_{l^+})\,\,
\epsilon^{\alpha \beta \sigma \rho}\,p_{1\alpha}\,p_{2\beta}\,p_{l^-
\sigma}\,p_{l^+ \rho}  \bigg],
\end{eqnarray}}
In the above, we have used
\be
F = \frac{m_Z}{s - m_Z^2} \left( \frac{e}{2\sin\theta_W\cos\theta_W}
		\right)^2 ,
\ee
the longitudinal polarizations $P_L$ and $\overline P_L$ of the electron
and positron, respectively, 
and the effective polarization
\be
P_L^{\rm eff} = \frac{P_L - \overline P_L}{1 - P_L \overline P_{L}}.
\ee

%\end{document}

\subsection{Distributions for transversely polarized beams}

For the transverse case, we take the $e^-$ polarization to be along the
$x$ axis 
and that of the $e^+$ to be antiparallel to that of the $e^-$. 
We define a four-vector $n^\mu=(0,1,0,0)$ and write the spin four-vector 
of the $e^-$ and $e^+$ as $n^{\mu}$ and $-n^{\mu}$ respectively. 
As before, the cross section consists of four
pieces coming from the square of the SM amplitude, proportional to
$\mathcal{F}^T_{\rm SM}$, and the interference
of the SM amplitude with the three contributions from $V_i$, $A_i$
($i=1,2,3$), respectively proportional to $\mathcal{F}^T_i$. The
expression for the cross section with transverse polarization $P_T$ for
the $e^-$ beam and $\overline P_T$ for the $e^+$ beam is
\begin{eqnarray}
\sigma_{\rm T}& =& \int \frac{d^3 p_{l^-}}{2p^0_{l^-}} \int \frac{d^3
p_{l^+}}{2p^0_{l^+}}
\left(\frac{e}{4 \sin \theta_W \cos \theta_W}\right)^2
\frac{1}{(q^2 -m_Z ^2)^2 +\Gamma_Z ^2 m_Z^2}
\nonumber \\
&&\times
\left[ \mathcal{F}^{\rm T}_{\rm SM} +
\mathcal{F}^{\rm T}_1+
\mathcal{F}^{\rm T}_2+
\mathcal{F}^{\rm T}_3 \right].
\end{eqnarray}
The expressions for the various $\mathcal{F}^{\rm T}$ are:
{\setlength\arraycolsep{2pt}
\begin{eqnarray}
\mathcal{F}^{\rm T}_{\rm SM} &=&4 F^2 \bigg[2\left\{ (g_V^2 +g_A^2)^2 -(g_V^4  -g_A^4)P_T \overline{P}_T\right\} \nonumber \\
&& \times \left\{(p_1 \cdot p_{l^-})(p_2 \cdot p_{l^+}) +(p_1 \cdot p_{l^+})(p_2 \cdot p_{l^-})\right\} \nonumber \\
&&+ s (g_V^4 -g_A^4)P_T \overline{P}_T \left\{2 (p_{l^-}\cdot n)(p_{l^+}\cdot n)+(p_{l^-} \cdot p_{l^+})\right\} \nonumber \\
&&-8 g_V^2 g_A^2 \left\{(p_1 \cdot p_{l^-})(p_2 \cdot p_{l^+}) -(p_1 \cdot p_{l^+})(p_2 \cdot p_{l^-})\right\} \bigg],
\end{eqnarray}
%%%%%%%%%%%%%%%%%%%%%%%
{\setlength\arraycolsep{2pt}
\begin{eqnarray}
\mathcal{F}^{\rm T}_{1} &=& \frac{8F}{M} \bigg[2(g_V^2
+g_A^2)\left\{g_V \textrm{Re}V_1\,(1-P_T\overline P_T) -g_A
\textrm{Re}A_1 \, (1 + P_T \overline P_T )\right\} \nonumber \\
&&\hskip -.8cm\times \left\{(p_1 \cdot p_{l^-})(p_2 \cdot p_{l^+}) +(p_1 \cdot p_{l^+})(p_2 \cdot p_{l^-})\right\} \nonumber \\
&&\hskip -.8cm+s(g_V^2 +g_A^2)(g_V \textrm{Re}V_1 +g_A \textrm{Re}A_1)P_T \overline{P}_T\left\{2 (p_{l^-}\cdot n)(p_{l^+}\cdot n)+(p_{l^-} \cdot p_{l^+})\right\} \nonumber \\
&&\hskip -.8cm-2 (g_V^2 +g_A^2)(g_V \textrm{Im}A_1 +g_A
\textrm{Im}V_1)P_T \overline{P}_T \left\{(p_{l^-}\cdot n)
p_{l^+}^{\beta}+(p_{l^+}\cdot n) p_{l^-}^{\beta}\right\}\nonumber \\
&&\hskip -.8cm\times \, \epsilon_{\alpha \beta \mu \nu}\,
n^{\alpha}\,p_{1}^{\nu}\,p_{2}^{\mu}\nonumber \\
&&\hskip -.8cm +4 g_V g_A (g_V \textrm{Re}A_1 -g_A \textrm{Re}V_1)
\left\{(p_1 \cdot p_{l^-})(p_2 \cdot p_{l^+})\! -\! (p_1 \cdot p_{l^+})(p_2 \cdot p_{l^-})\right\}\bigg],
\end{eqnarray}
%%%%%%%%%%%%%%%%%%%%%%%%%%%%%%%%%
{\setlength\arraycolsep{2pt}
\begin{eqnarray}
\mathcal{F}^{\rm T}_{2} &=& -\frac{8F}{M^3} 
\bigg[(g_V^2 +g_A^2)
\left\{g_V \textrm{Re}V_2\, (1 - P_T \overline P_T) -g_A \textrm{Re}A_2
\, (1 + P_T \overline P_T) \right\} 
\nonumber \\
&&\hskip -.0cm \times \left\{(p_1\cdot q)[(p_2\cdot p_{l^-})(P\cdot p_{l^+})
+(p_2\cdot p_{l^+})(P\cdot p_{l^-}) 
-(p_{l^-}\cdot p_{l^+})\left(s/2 \right)]\right. \nonumber \\
&&\hskip -.0cm +\left. (p_2 \cdot q)[(p_1\cdot p_{l^-})(P\cdot p_{l^+})
+(p_1\cdot p_{l^+})(P\cdot p_{l^-}  )
 -(p_{l^-}\cdot p_{l^+})\left(s/2 \right)]\right\}\nonumber \\
&&\hskip -.0cm -2g_V g_A \left\{(g_V \textrm{Im}V_2 -g_A \textrm{Im}A_2)-(g_V \textrm{Im}V_2 +g_A \textrm{Im}A_2)P_T \overline{P}_T\right\} \nonumber \\
&&\times (p_2-p_1)\cdot q \,\,\epsilon_{\alpha \beta \sigma \rho}\,p_1^{\alpha}\,p_2^{\beta}\,p_{l^-}^{\sigma}\,p_{l^+}^{\rho} \nonumber \\
&&+s\,(g_V^2+g_A^2)(g_V \textrm{Re}V_2 +g_A \textrm{Re}A_2)P_T \overline{P}_T
(q\cdot n)\nonumber \\
&&\left\{(P\cdot p_{l^+}) (p_{l^-}\cdot n)+ (P\cdot p_{l^-}) (p_{l^+}\cdot n)\right\}\nonumber \\
&&+2s\,g_V g_A (g_V \textrm{Im}V_2 +g_A \textrm{Im}A_2)P_T \overline{P}_T
(q\cdot n)\,\,\epsilon_{\mu \nu \sigma
\rho}\,P^{\mu}n^{\nu}p_{l^-}^{\sigma}p_{l^+}^{\rho} \nonumber \\
&&-(g_V^2+g_A^2)(g_V \textrm{Im}A_2 -g_A
\textrm{Im}V_2)P\cdot(p_{l^+}-p_{l^-})\,\epsilon_{\mu \nu \sigma
\rho}\,p_{1}^{\mu}p_{2}^{\nu}p_{l^-}^{ \sigma}p_{l^+}^{ \rho} \nonumber \\
&&+2 g_V g_A (g_V \textrm{Re}A_2-g_A \textrm{Re}V_2)\nonumber \\
&&\{P\cdot q[(p_1\cdot p_{l^-})(p_2\cdot p_{l^+})
-(p_2\cdot p_{l^-})(p_1\cdot p_{l^+})]\nonumber \\
&&+ (s/2) (p_{l^-}\cdot p_{l^+})\,(p_1-p_2)\cdot (p_{l^+}-p_{l^-})\}\nonumber \\
&&+(g_V^2+g_A^2)(g_V \textrm{Im}A_2+g_A\textrm{Im}V_2)P_{T}\overline{P}_T 
\nonumber \\
&& \times\,\, \epsilon_{\mu \nu \alpha
\sigma}\,p_{1}^{\mu}\,p_{2}^{\nu}\,n^{\alpha}\,q^{\sigma}\, \{P\cdot p_{l^+} (p_{l^-}\cdot n) +P\cdot p_{l^-} (p_{l^+}\cdot n)\} \nonumber \\
&&+2g_V g_A (g_V \textrm{Re}A_2+g_A \textrm{Re}V_2)P_{T}\overline{P}_T\,
\nonumber \\
%\epsilon^{\mu \nu \alpha \sigma}\,p_{1 \mu}\,p_{2 \nu}\,n_{\alpha}\,q_{\sigma}\nonumber \\
%&&\times  \epsilon^{\mu' \nu' \alpha' \sigma'}\,n_{\mu'}\,P_{\nu'}\,p_{l^- \alpha'}\,p_{l^+ \sigma'}\nonumber \\
&&  [ - q\cdot P ( p_1\cdot p_{l^+}\, p_2 \cdot p_{l^-} 
 - p_1 \cdot p_{l^-}\,
p_2 \cdot  p_{l^+}) \nonumber \\
&&  + s\, q\cdot n ( p_{l^+}\cdot (p_1 -  p_2)\,
p_{l^-}\cdot n -  p_{l^-} \cdot (p_1 -  p_2)\,  p_{l^+}\cdot n ) \nonumber
\\
&& + (s/2)(p_{l^-} \cdot p_{l^-})( p_{l^+}- p_{l^-}) \cdot (p_1 -  p_2)
 ] \nonumber \\
&&+(g_V^2+g_A^2)(g_V\textrm{Im}A_2
+g_A\textrm{Im}V_2)P_{T}\overline{P}_T \{P\cdot
p_{l^-}\,p_{l^+}^{\sigma}+P\cdot p_{l^+}\,p_{l^-}^{\sigma}\} \nonumber \\
&&\times (q\cdot n)\,\epsilon_{\mu \nu \alpha
\sigma}\,p_{1}^{\mu}\,p_{2}^{\nu}\,n^{\alpha}
%\epsilon^{\mu \nu \alpha \beta}\,\epsilon_{\alpha' \beta \sigma' \rho'}(q\cdot n)\nonumber \\
%&&\times \,p_{1\mu}\,p_{2\nu}\,n_{\alpha}P^{\alpha'}\,p_{l^-}^{\sigma'}\,p_{l^+}^{\rho'}
\bigg],
\end{eqnarray}
and
%%%%%%%%%%%%%%%%%%%%%%%%%%%%%
{\setlength\arraycolsep{2pt}
\begin{eqnarray}
\mathcal{F}^{\rm T}_{3} &=& \frac{8F}{M^3} \bigg[-(g_V^2 +g_A^2)\left\{g_V \textrm{Im}V_3(1-P_T \overline{P}_T)-g_A \textrm{Im}A_3(1+P_T \overline{P}_T)\right\}\nonumber \\
&&\times \left\{
(p_2-p_1)\cdot p_{l^+}[(p_1 \cdot q)(p_2\cdot p_{l^-})
+(p_2 \cdot q)(p_1\cdot p_{l^-})- (s/2) (p_{l^-}\cdot p_{l^+})]\right. \nonumber \\
&&\left.   + 
\;\,\, (p_2-p_1)\cdot p_{l^-}[(p_1 \cdot q)(p_2\cdot p_{l^+})
+(p_2 \cdot q)(p_1\cdot p_{l^+})- (s/2)( p_{l^-}\cdot p_{l^+})]
\right\}\nonumber \\
%&&\times \left\{(p_1 \cdot q)[(p_2\cdot p_{l^-})(p_2-p_1)\cdot p_{l^+}
%+(p_2\cdot p_{l^+})(p_2-p_1)\cdot p_{l^-} 
%+\frac{s}{2}(p_{l^-}\cdot p_{l^+})]\right.  \nonumber \\
%&&\left. +(p_2 \cdot q)[(p_1\cdot p_{l^-})(p_2-p_1)\cdot p_{l^+} 
%+(p_1\cdot p_{l^+})(p_2-p_1)\cdot p_{l^-} 
%-\frac{s}{2}(p_{l^-}\cdot p_{l^+})]\right\}\nonumber \\
&&-2\,g_V g_A\left\{g_V \textrm{Re}V_3(1-P_T \overline{P}_T)-g_A \textrm{Re}A_3(1+P_T \overline{P}_T)\right\}\nonumber \\
&&\times (P\cdot q)\, \epsilon_{\mu \nu \sigma
\rho}\,p_{1}^{\mu}\,p_{2}^{\nu}\,p_{l^-}^{\sigma}\,p_{l^+}^{\rho} \nonumber \\
&&+(g_V^2+g_A^2)\left\{g_A \textrm{Re}V_3-g_V\textrm{Re}A_3\right\}\, 
\epsilon_{\mu  \nu  \rho  \sigma}\,
p_{1}^{\mu}\,p_{2}^{\nu}p_{l^-}^{\sigma}p_{l^+}^{\rho}\,(p_2-p_1)\cdot (p_{l^-} -
p_{l^+})\nonumber \\
%\epsilon^{\mu \, \nu \, \rho \, \sigma}\, p_{1\mu}\,p_{2\nu}\nonumber \\
%&&\times [((p_2-p_1)\cdot p_{l^-})\,p_{l^- \rho}\,p_{l^+\sigma}+((p_2-p_1)\cdot p_{l^+})\,p_{l^- \sigma}\,p_{l^+\rho}]\nonumber \\
%&&-2\,g_V g_A \left\{g_V \textrm{Im}A_3-g_A \textrm{Im}V_3\right\}\,\epsilon^{\mu \nu \beta \sigma}\, \epsilon_{\alpha \beta \delta \rho}\,p_{1\mu}\,p_{2\nu}\,q_{\sigma}\,p_{l^-}^{\delta}\,p_{l^+}^{\rho}\,(p_2-p_1)^{\alpha} \nonumber \\
&&-2\,g_V g_A \left\{g_V \textrm{Im}A_3-g_A
\textrm{Im}V_3\right\}\,\left[ (s/2)\, (p_{l^-}\cdot p_{l^+})\, P\cdot
(p_{l^-} -p_{l^+}) \right. \nonumber \\
&& \left. + q\cdot (p_2-p_1) \, \left\{ p_1\cdot p_{l^-}
p_2\cdot p_{l^+} -p_1\cdot p_{l^+} p_2\cdot p_{l^-} \right\} \right]
\nonumber \\
&&-s \,(g_V^2+g_A^2)(g_V\textrm{Im}V_3+g_A \textrm{Im}A_3)P_{T}\overline{P}_T
\nonumber \\
&&\times (q\cdot n)\{(p_2-p_1)\cdot p_{l^+}p_{l^-}\cdot n
+(p_2-p_1)\cdot p_{l^-}p_{l^+}\cdot n\} \nonumber \\
&&-2s\,g_Vg_A(g_V\textrm{Re}V_3+g_A\textrm{Re}A_3)P_{T}\overline{P}_T(q\cdot
n)\epsilon_{\alpha \beta \sigma
\rho}\,n^{\alpha}\,(p_2-p_1)^{\beta}\,p_{l^-}^{\sigma}\,p_{l^+}^{\rho}
\nonumber \\
%&&+(g_V^2+g_A^2)\left\{(g_V\textrm{Re}A_3 +g_A\textrm{Re}V_3)P_{T}\overline{P}_T\right\} \,\epsilon^{\mu \nu \sigma \rho}\,p_{1\mu}\,p_{2\nu}\,n_{\sigma}\,q_{\rho} \nonumber \\
%&&\times \{((p_2-p_1)\cdot p_{l^+})(p_{l^-}\cdot n)+((p_2-p_1)\cdot p_{l^-})(p_{l^+}\cdot n)\} \nonumber \\
&&+(g_V^2+g_A^2)(g_V\textrm{Re}A_3+g_A \textrm{Re}V_3)P_{T}\overline{P}_T 
\epsilon_{\mu \nu \sigma \rho}\,p_{1}^{\mu}\,p_{2}^{\nu} \,n^{\sigma}
\nonumber \\
&&\times \left\{ q^{\rho}
[(p_2-p_1)\cdot p_{l^+}p_{l^-}\cdot n
+(p_2-p_1)\cdot p_{l^-}p_{l^+}\cdot n]\right. \nonumber \\
%&&+(g_V^2+g_A^2)(g_V\textrm{Re}A_3+g_A\textrm{Re}V_3)P_{T}\overline{P}_T 
%\epsilon^{\mu \nu \sigma \rho}\,p_{1\mu}\,p_{2\nu} \nonumber \\
&&\left. + (q\cdot n)
 [(p_2-p_1)\cdot p_{l^-}\,p_{l^+}^{ \rho}+(p_2-p_1)\cdot
p_{l^+}\,p_{l^-}^{ \rho}]
\right\} \nonumber \\
&&+2 g_Vg_A(g_V\textrm{Im}A_3+g_A\textrm{Im}V_3)P_{T}\overline{P}_T\nonumber \\
%&&\times \epsilon^{\mu \nu \sigma \rho}\,p_{1\mu}\,p_{2\nu}\, n_{\sigma}\, q_{\rho}\,\epsilon_{\alpha \beta \sigma' \rho'}\,n^{\alpha}\,(p_2-p_1)^{\beta}\,p_{l^-}^{\sigma'}\,p_{l^+}^{\rho'}\nonumber \\
%&&\times \left[ p_1\cdot p_2 n\cdot P n\cdot ( p_{l^-}-p_{l^+}) p_{l^-}\cdot
%p_{l^+} - p_1\cdot p_{l^-} n\cdot p_2 n\cdot (p_2-p_1) p_{l^-}\cdot
%p_{l^+} +  \right] \nonumber \\
&&\times \left\{ \left[ q\cdot (p_1-p_2)\, (p_1\cdot p_{l^-}\, p_2 \cdot
p_{l^+} - p_1\cdot p_{l^+}\, p_2 \cdot p_{l^-} )
 \right. \right. \nonumber \\
&&\left. \left. - (s/2)\, p_{l^-}\cdot p_{l^+}\, P\cdot (p_{l^-} 
- p_{l^+} )
 \right]\right. \nonumber \\
%&&-2 g_Vg_A(g_V\textrm{Im}A_3+g_A\textrm{Im}V_3)P_{T}\overline{P}_T\nonumber \\
%&&\times (q\cdot n)\,\epsilon^{\mu \nu \sigma \beta}\,p_{1\mu}\,p_{2\nu}\, n_{\sigma}\,\epsilon_{\alpha \sigma' \rho' \beta}\,(p_2-p_1)^{\alpha}\,p_{l^-}^{\sigma'}\,p_{l^+}^{\rho'}\bigg]
&&\left. +s\, q\cdot n\, ( P\cdot p_{l^+} n \cdot p_{l^-} -
P\cdot p_{l^-} n \cdot p_{l^+} ) \right\}
\bigg].
\end{eqnarray}

\section{Observables}

\begin{table}[!htb]
\centering
\begin{tabular}{|c|c|c|c|}
\hline
Symbol & Observable & CP & T \\
\hline
$O_1$ &  $(p_1 - p_2)\cdot q$ & $-$ & $+$ \\
$O_2$ & $(\vec p_{l^-} \times \vec p_{l^+})_z$ & $-$ & $-$ \\
$O_3$ & $(p_1 - p_2)\cdot q\,(\vec p_{l^-} \times \vec p_{l^+})_z$& $+$
& $-$ \\
$O_4$ &$(p_1 - p_2)\cdot  ( p_{l^-} - p_{l^+})$ & $+$ & $+$ \\
$O_5$ &  $(p_1 - p_2)\cdot  ( p_{l^-} - p_{l^+})\, (\vec p_{l^-} \times
\vec p_{l^+})_z$ & $-$ & $-$ \\
$O_6$ &   $q_x\, q_y\, q_z$ & $-$ & $-$ \\
$O_7$ &  $( q_x^2 -   q_y^2)\, q_z$ & $-$ & $+$ \\
$O_8$ & $(\vec p_{l^-} - \vec p_{l^+})_x\, (\vec p_{l^-} - \vec
p_{l^+})_y\, q_z$ & $-$ & $-$ \\
$O_9$ & $q_x\, q_y\, (\vec p_{l^-} - \vec p_{l^+})_z$ & $+$ & $-$ \\
$O_{10}$ &  $P\cdot  ( p_{l^-} - p_{l^+}) $ & $-$ & $+$ \\
\hline
\end{tabular}
\caption{Observables whose expectation values can be used to constrain
the form factors, and their CP and T properties}\label{obs}
\end{table}

After having obtained analytical expressions for the differential cross 
section, we choose observables whose
expectation values can be
used to constrain the form factors. We have chosen observables which
have well defined CP properties. Thus, the expectation values of 
observables which are even
under CP get contributions from $V_1$, $V_2$, $A_1$ and $A_2$, with our
assumption that the form factors have only $s$ dependence, and no
dependence on $t$ (or that they are even functions of $t-u$). On the
other hand, expectation values of observables which are odd under CP 
get contribution only from $V_3$ and $A_3$. Moreover, it is important to
note the behaviour of the observables under na\" ive time reversal T.
The CPT theorem implies that observables which are CP even and T even would get contributions from
only real parts of the form factors, as also those which are CP odd and
T odd \cite{sdrCP}. On the other hand, observables which are 
CP even and T odd, or CP odd and T even, are CPT odd. They therefore
require the presence of an absorptive part in the amplitude, and hence are
proportional to the imaginary parts of form factors \cite{sdrCP}.  

The observables we choose are listed in Table~\ref{obs}, together with
their CP and T properties.

We have evaluated the expectation values of the observables $O_i$
($i=1, 2,\dots , 10$) for unpolarized as well as polarized beams,
choosing the $e^-$ polarization to be 0.8 and $e^+$ polarization to be
$\pm 0.6$. The relevant phase-space integrals have been done
numerically. In calculating the expectation values, we use the
expressions for the differential cross sections given above to leading
order in the new-physics contact interactions in the formula
\be\label{expvalue} 
\langle O_i \rangle = \frac{1}{\sigma_{\rm SM}}\, \int O_i\frac{
d\sigma}{ d^3 p_{l^-}d^3p_{l^+}}\, d^3 p_{l^-}d^3p_{l^+}.
\ee
Since the expectation
values we concentrate are vanishing in SM, for
consistency, we need use only the SM cross section in the denominator in
eq.~(\ref{expvalue}). 
We assume a cut-off of $\theta_0$ in the forward and
backward directions on the azimuthal angles of both leptons. Such a
cut-off is an experimental requirement, so as to avoid the beam pipe.
However, we also explore the possibility that a suitable choice of
$\theta_0$ can optimize the sensitivity. 
We treat both cases of longitudinal as well as transverse
polarization. 
We take one form factor to be nonzero at a time.
The expectation value of each observable for a given nonzero form factor
is compared with the standard deviation of the observable in 
SM. This allows us to determine a limit that nonobservance of the
expectation value can place on the corresponding form factor. Thus, the
90\% CL (confidence level) limit $F_{\rm lim}$ 
on a form factor $F$ is determined by the expression
\be\label{eq:limits}
F_{\rm lim} = 1.64 \frac{\sqrt{\langle O^2 \rangle/\sigma _{\rm SM}}}
{\langle O \rangle_1 \sqrt{L}},
\ee
where $\sigma$ is the SM cross section, $\langle O \rangle_1$ the
expectation value of the observable $O$ for unit value of the form
factor, and $L$ is the integrated luminosity.

\section{Numerical results}

\begin{figure}[!ht]
\centering
\epsfig{height=7cm,file=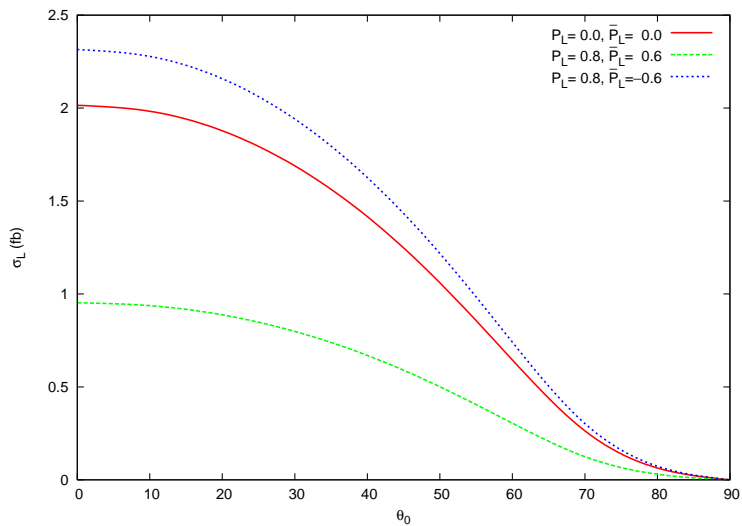}
\caption{The cross section in fb for SM as a function of the cut-off angle
$\theta_0$ for unpolarized beams and for longitudinally polarized beams.}
\label{fig:cslong}\end{figure}
%\begin{figure}[!hbt]
%\centering
%\epsfig{height=7cm,file=diffcrosstrans.eps}
%\caption{The cross section in fb for SM as a function of the cut-off angle
%$\theta_0$ for transversely polarized beams.}
%\label{fig:cstrans}\end{figure}

We now present out numerical results for the correlations and the
limits that can be obtained from the form factors. We assume a c.m.
energy of 500~GeV for the linear collider, and polarizations $\pm 0.8$
and $\pm 0.6$ respectively 
for the electron and positron beams.  We have chosen the
value of the arbitrary mass scale $M$ to be 100~GeV.
We vary the cut-off $\theta_0$.

First of all, we present the SM cross sections in the cases of
unpolarized beams or longitudinal polarized beams in Fig.
\ref{fig:cslong} as a function of the cut-off angle $\theta_0$. 
In case of transverse beam polarization, the cross section when
integrated over the azimuthal angle $\phi$ reduces to the unpolarized
cross section \cite{hikasa}, and hence is not shown separately.

Next, we present expectation values of some of the observables as
functions of the cut-off $\theta_0$ for the purpose of illustration. The
limits obtainable on the form factors, however, are presented for 
all observables  in tables.

Figs. \ref{fig:1l10}-\ref{fig:10l12} show some sample correlations for
unpolarized and longitudinally polarized beams.
\begin{figure}[!t]
\centering
\epsfig{height=6.5cm,file=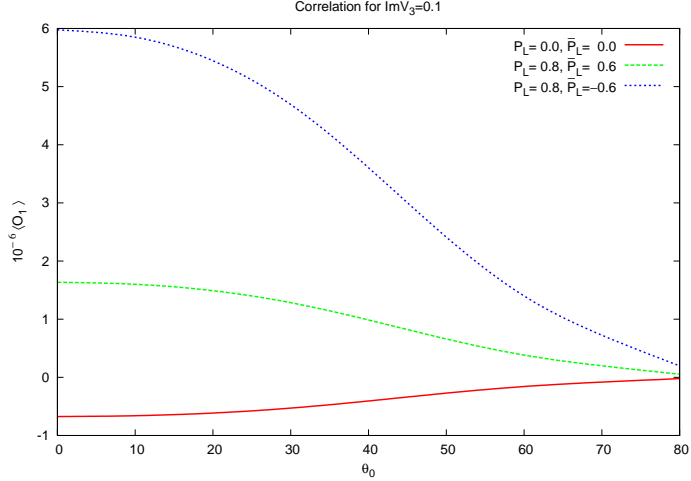}
\caption{The expectation value of $O_1$ (scaled down by an appropriate
factor) in GeV$^2$ with longitudinally
polarized beams for Im~$V_3=0.1$. The remaining form factors are
zero.}\label{fig:1l10}
\end{figure}
\begin{figure}[!hb]
\centering
\epsfig{height=6.5cm,file=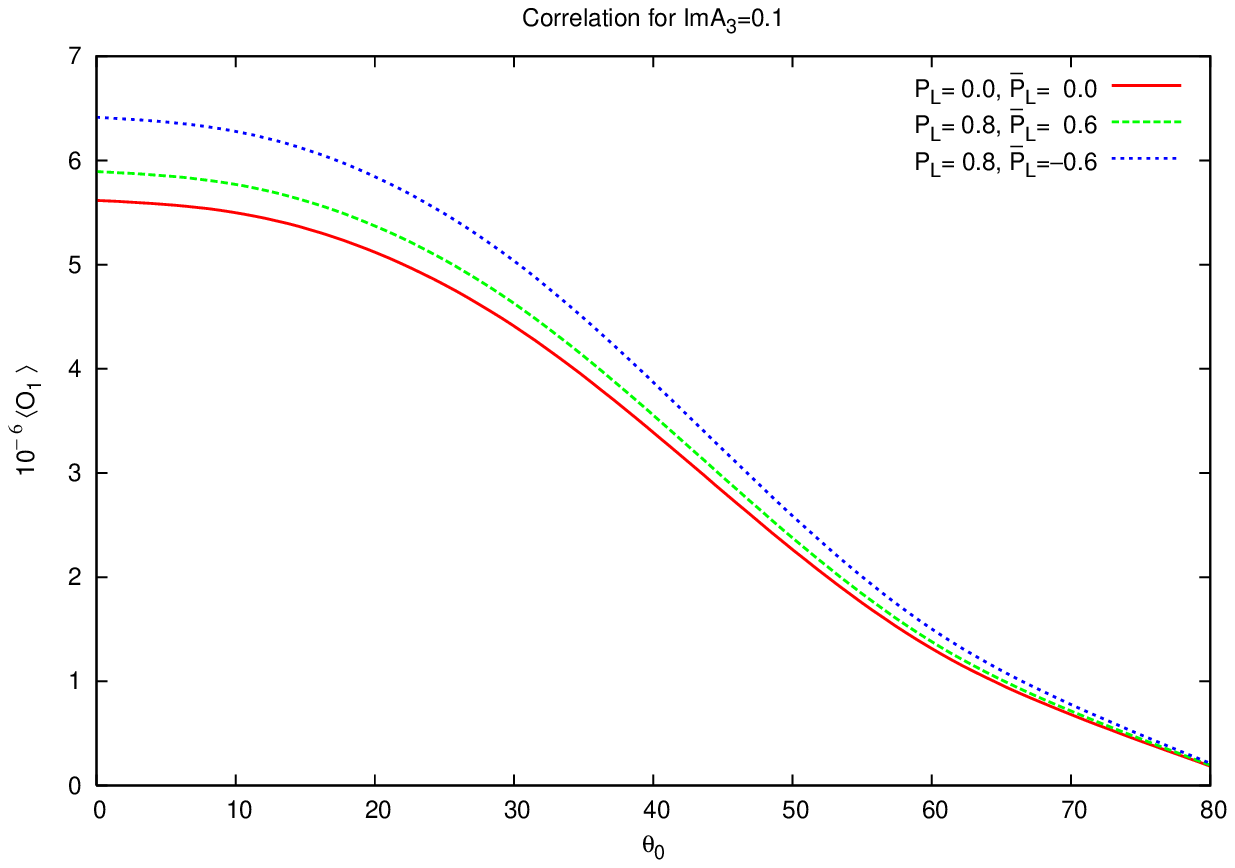}
%\centerline{ \tiny (b)}
\caption{The expectation value of $O_1$ (scaled down by an appropriate
factor) in GeV$^2$ with longitudinally
polarized beams for 
Im~$A_3=0.1$ and the remaining form factors
zero.}\label{fig:1l12}
\end{figure}

\begin{figure}[!t]
\centering
\epsfig{height=6.5cm,file=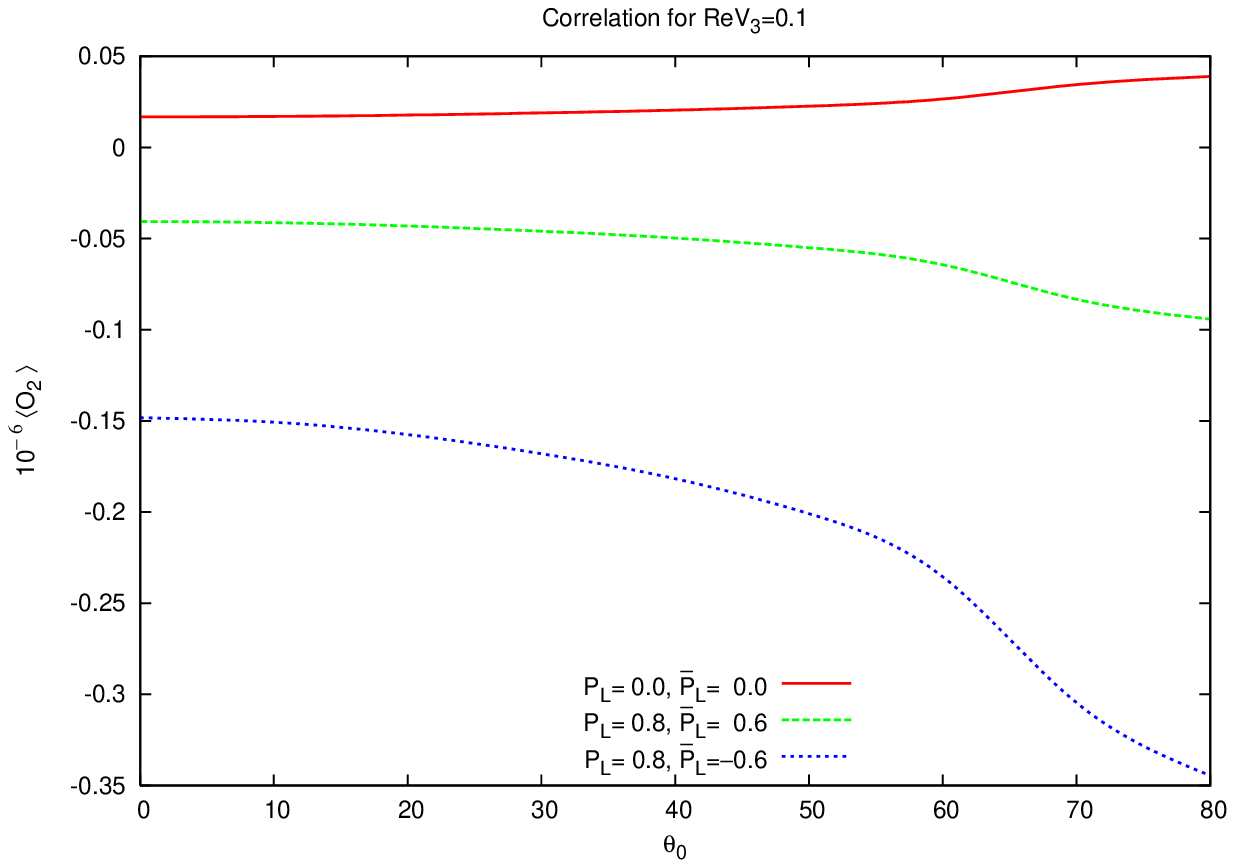}
\caption{The expectation value of $O_2$ (scaled down by an appropriate
factor) in GeV$^2$ with longitudinally
polarized beams with Re~$V_3=0.1$. The remaining form factors are
zero.}\label{fig:2l9}
\end{figure}
\begin{figure}[!ht]
\centering
\epsfig{height=6.5cm,file=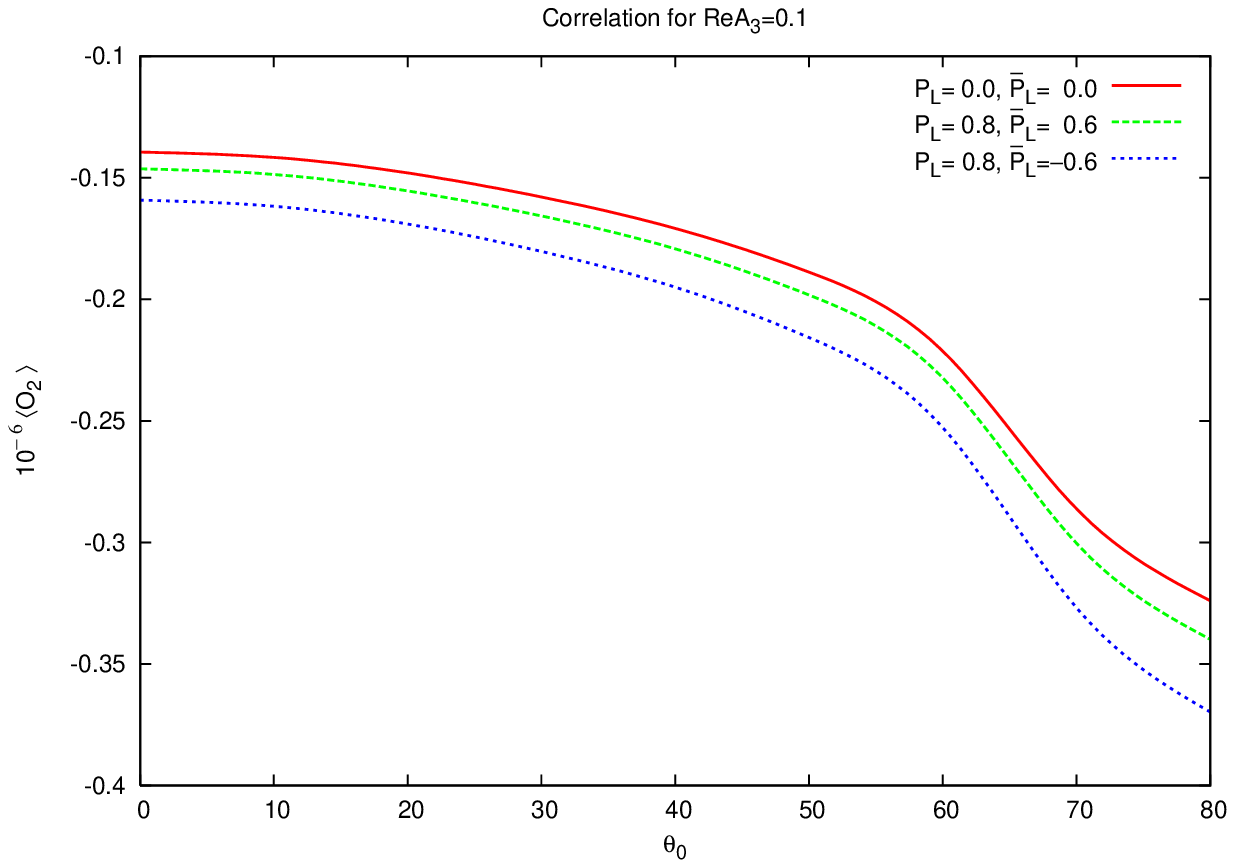}
\caption{The expectation value of $O_2$ (scaled down by an appropriate
factor) in GeV$^2$ with longitudinally
polarized beams with Re~$A_3=0.1$. The remaining form factors are
zero.}\label{fig:2l11}
\end{figure}

\begin{figure}[!ht]
\centering
\epsfig{height=6.5cm,file=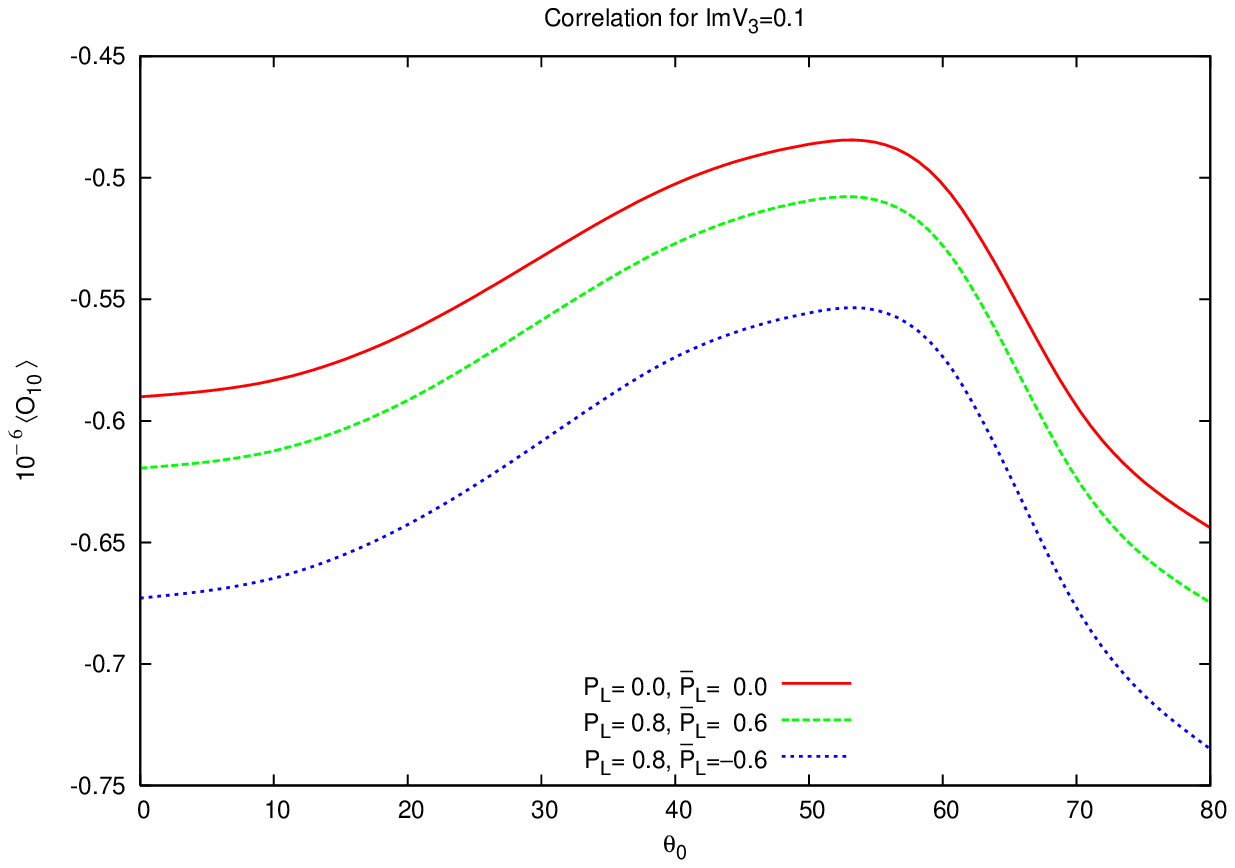}
\caption{The expectation value of $O_{10}$ (scaled down by an
appropriate factor) in GeV$^2$ with longitudinally
polarized beams with Im~$V_3=0.1$. The remaining form factors are
zero.}\label{fig:10l10}
\end{figure}
\begin{figure}[!ht]
\centering
\epsfig{height=6.5cm,file=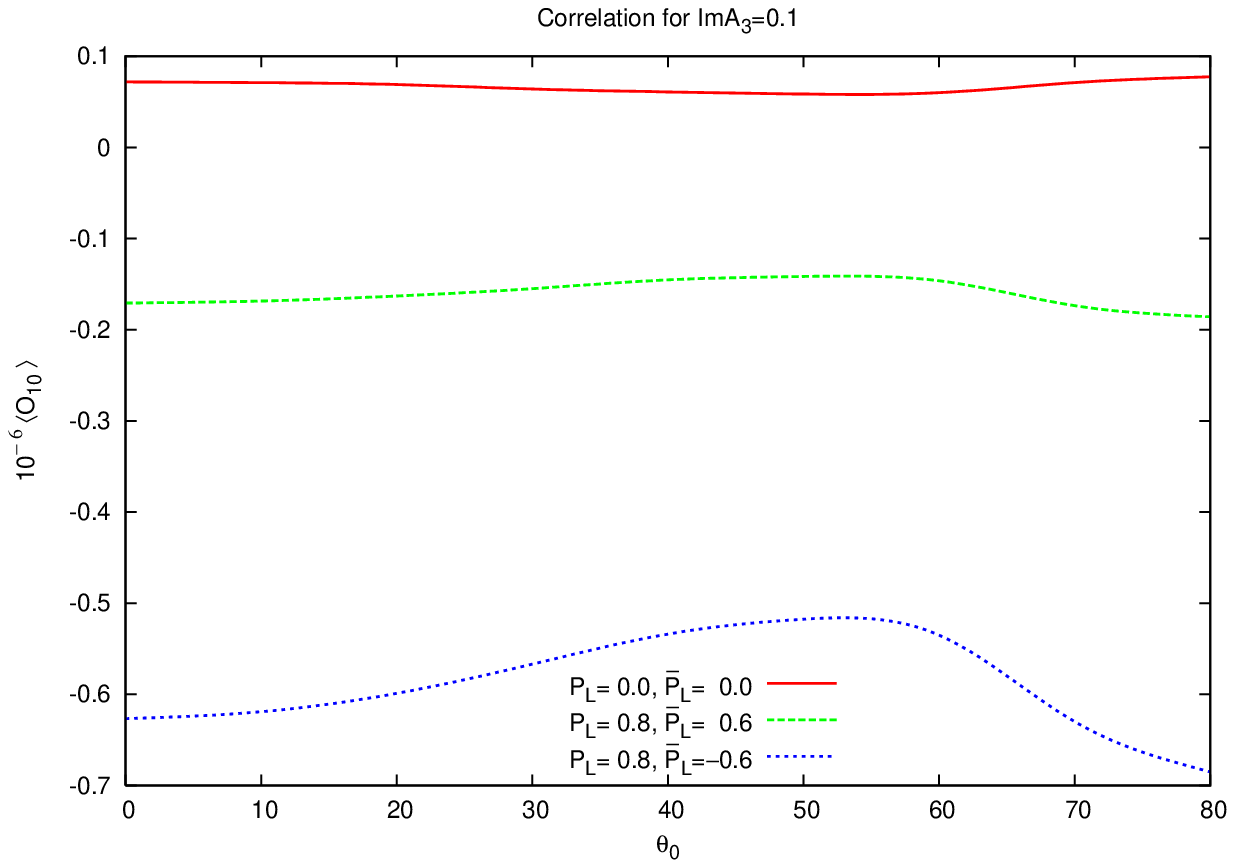}
\caption{The expectation value of $O_{10}$ (scaled down by an
appropriate factor) in GeV$^2$ with longitudinally
polarized beams with Im~$A_3=0.1$. The remaining form factors are
zero.}\label{fig:10l12}
\end{figure}
Fig. \ref{fig:1l10} shows the correlation for $O_1$ with unpolarized and 
longitudinally
polarized beams, when Im~$V_3=0.1$, and the rest of the form factors are
zero. Fig. \ref{fig:1l12}  is the corresponding figure for Im~$A_3=0.1$,
and the remaining form factors zero. 

Similarly, Figs. \ref{fig:2l9},
\ref{fig:2l11} give the correlations for $O_2$ for the two cases when
single form factors Re~$V_3=0.1$ and Re~$A_3=0.1$ are respectively
nonzero.

Finally, Figs. \ref{fig:10l10} and \ref{fig:10l12} give the correlations
for $O_{10}$ for the two cases with Im~$V_3=0.1$ and Im~$A_3=0.1$,
respectively. 

%\begin{figure}[!htb]
%\centering
%\epsfig{height=7cm,file=2tj9.eps}
%\caption{The expectation value of $O_2$ with transversely 
%polarized beams with Re~$V_3=0.1$. The remaining form factors are
%zero.}\label{fig:2t9}
%\end{figure}
%\begin{figure}[!htb]
%\centering
%\epsfig{height=7cm,file=2tj11.eps}
%\caption{The expectation value of $O_2$ with transversely 
%polarized beams with Re~$A_3=0.1$. The remaining form factors are
%zero.}\label{fig:2t11}
%\end{figure}
\begin{figure}[!htb]
\centering
\epsfig{height=6.5cm,file=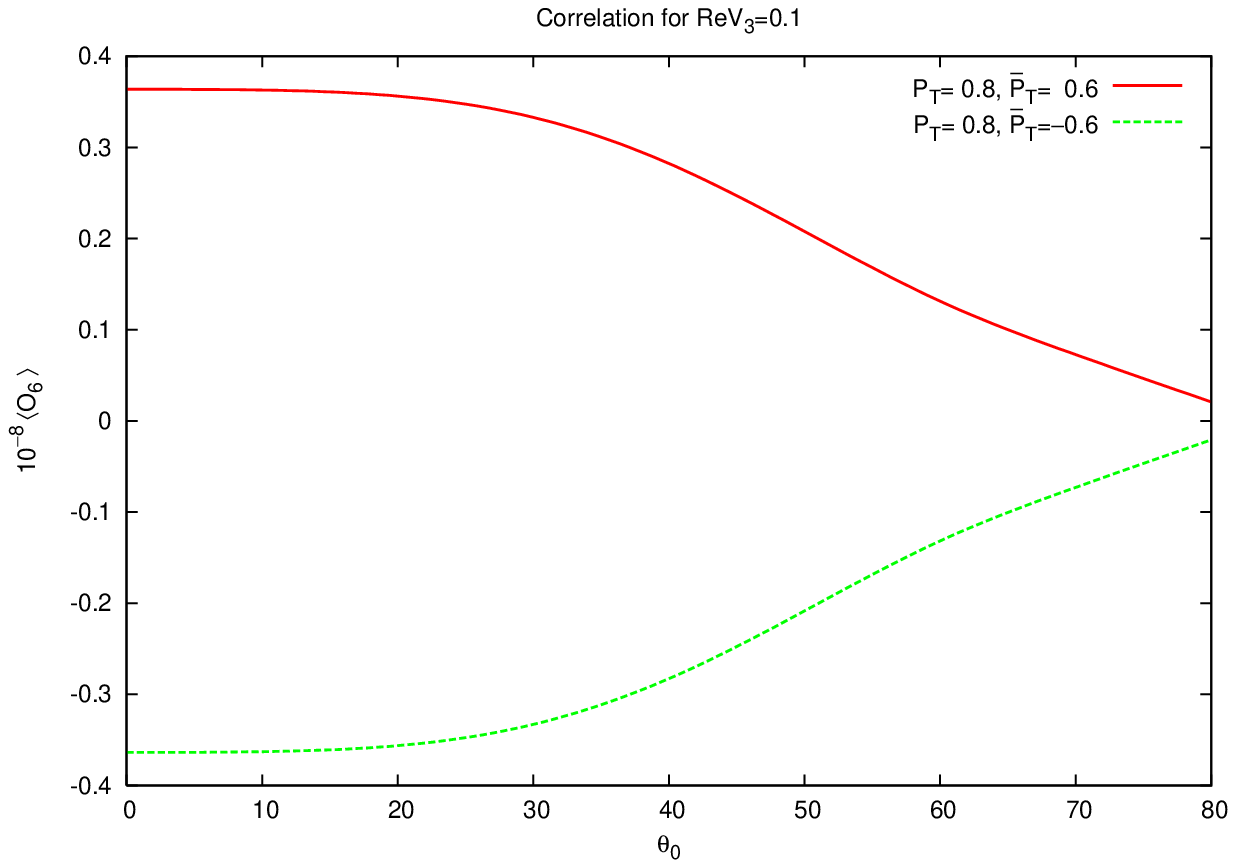}
\caption{The expectation value of $O_{6}$ (scaled down by an appropriate
factor) in GeV$^3$ with transversely 
polarized beams with Re~$V_3=0.1$. The remaining form factors are
zero.}\label{fig:6t9}
\end{figure}
\begin{figure}[!htb]
\centering
\epsfig{height=6.5cm,file=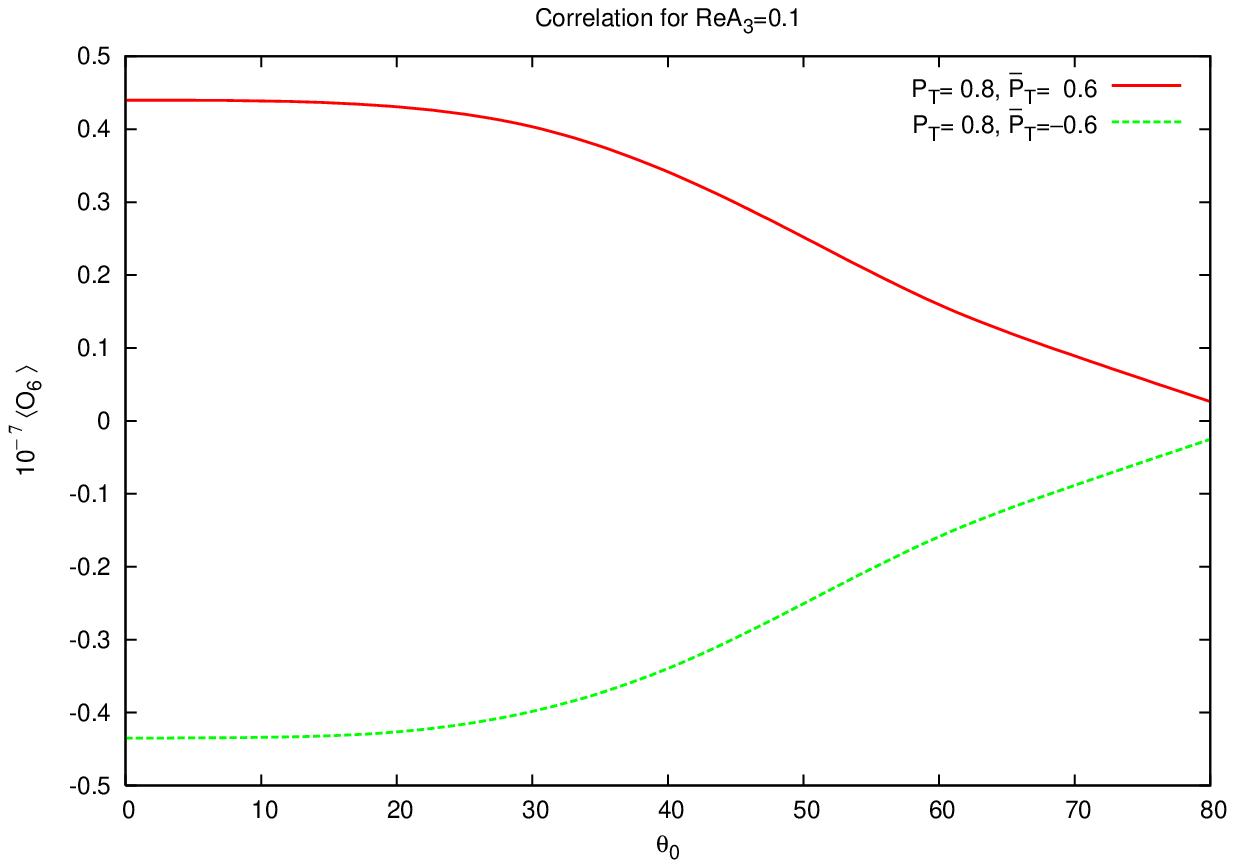}
\caption{The expectation value of $O_{6}$ (scaled down by an appropriate
factor) in GeV$^3$ with transversely 
polarized beams with Re~$A_3=0.1$. The remaining form factors are
zero.}\label{fig:6t11}
\end{figure}
\begin{figure}[!htb]
\centering
\epsfig{height=6.5cm,file=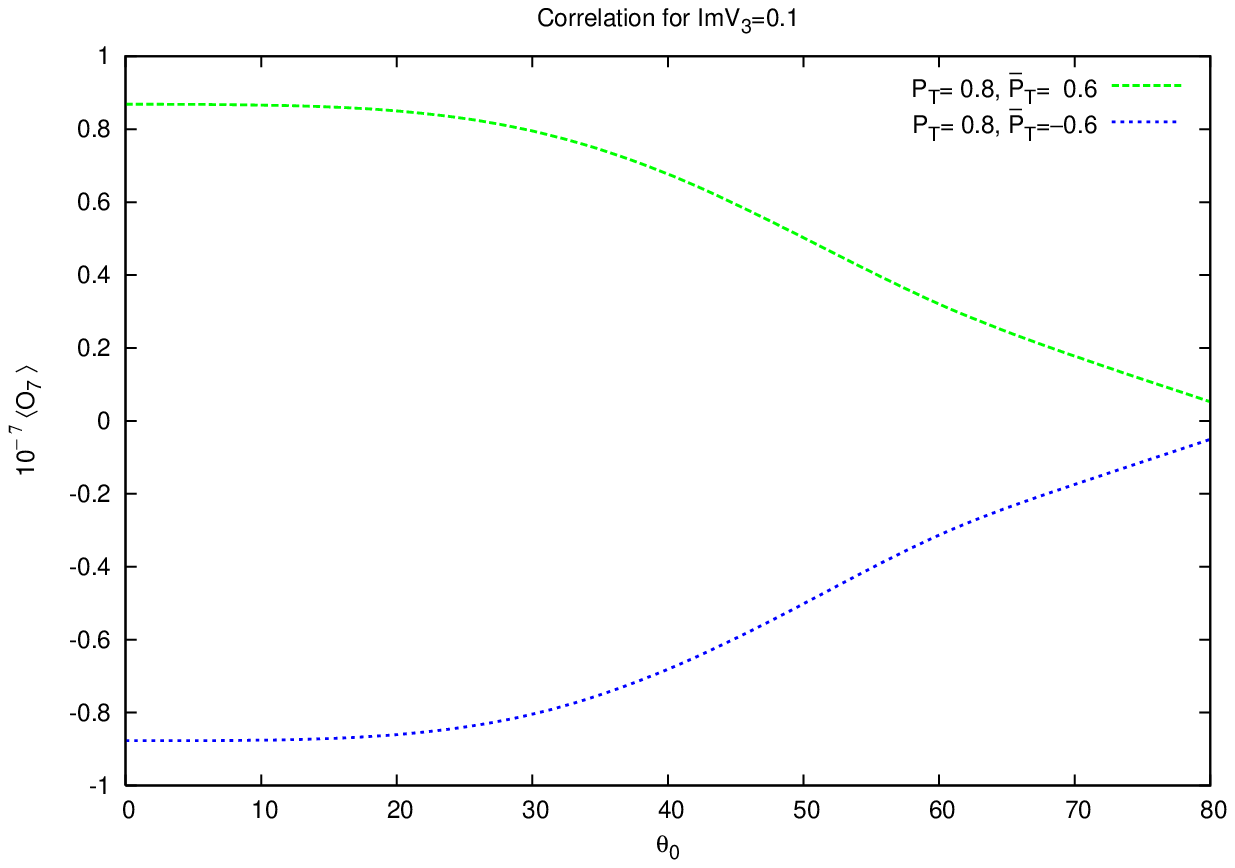}
\caption{The expectation value of $O_{7}$ (scaled down by an appropriate
factor) in GeV$^3$ with transversely 
polarized beams with Im~$V_3=0.1$. The remaining form factors are
zero.}\label{fig:7t10}
\end{figure}
\begin{figure}[!htb]
\centering
\epsfig{height=6.5cm,file=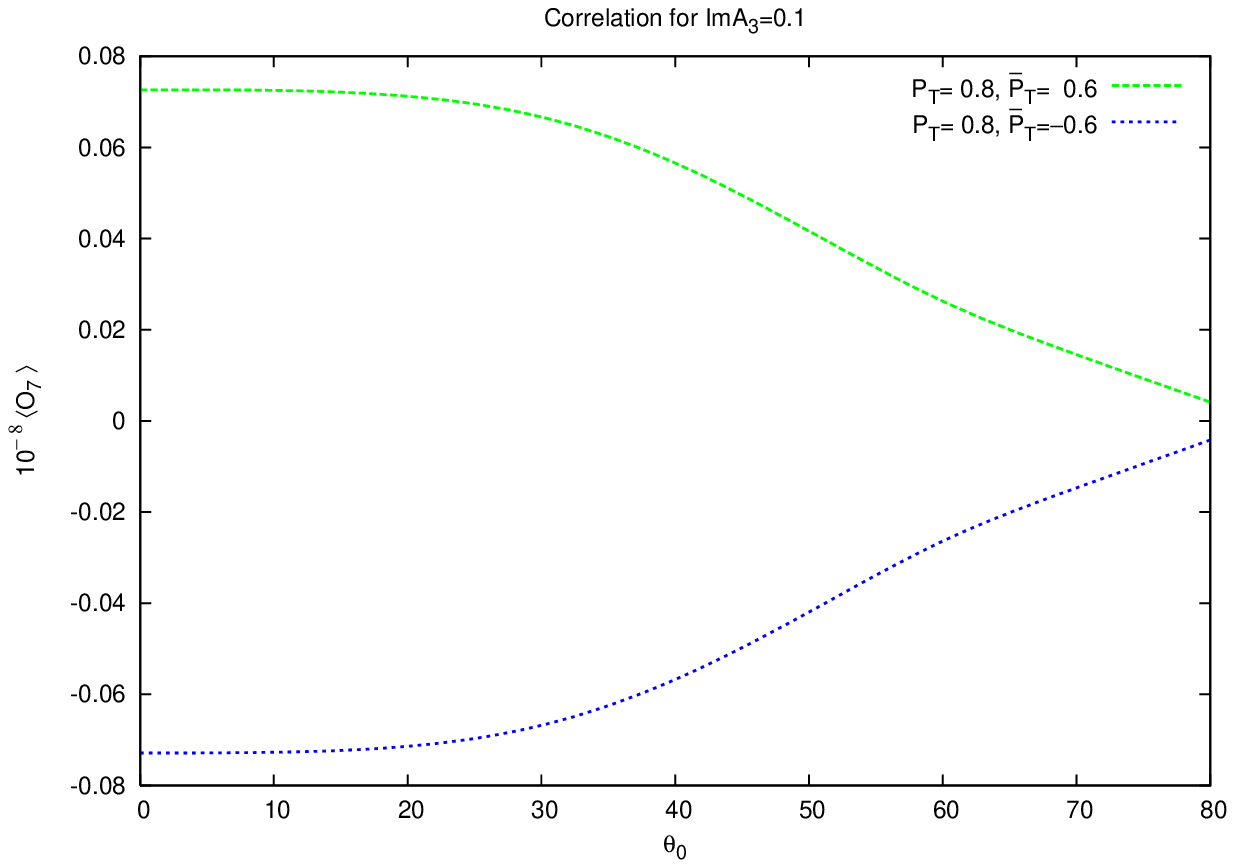}
\caption{The expectation value of $O_{7}$ (scaled down by an appropriate
factor) in GeV$^3$ with transversely 
polarized beams with Im~$A_3=0.1$. The remaining form factors are
zero.}\label{fig:7t12}
\end{figure}
\begin{figure}[!htb]
\centering
\epsfig{height=6.5cm,file=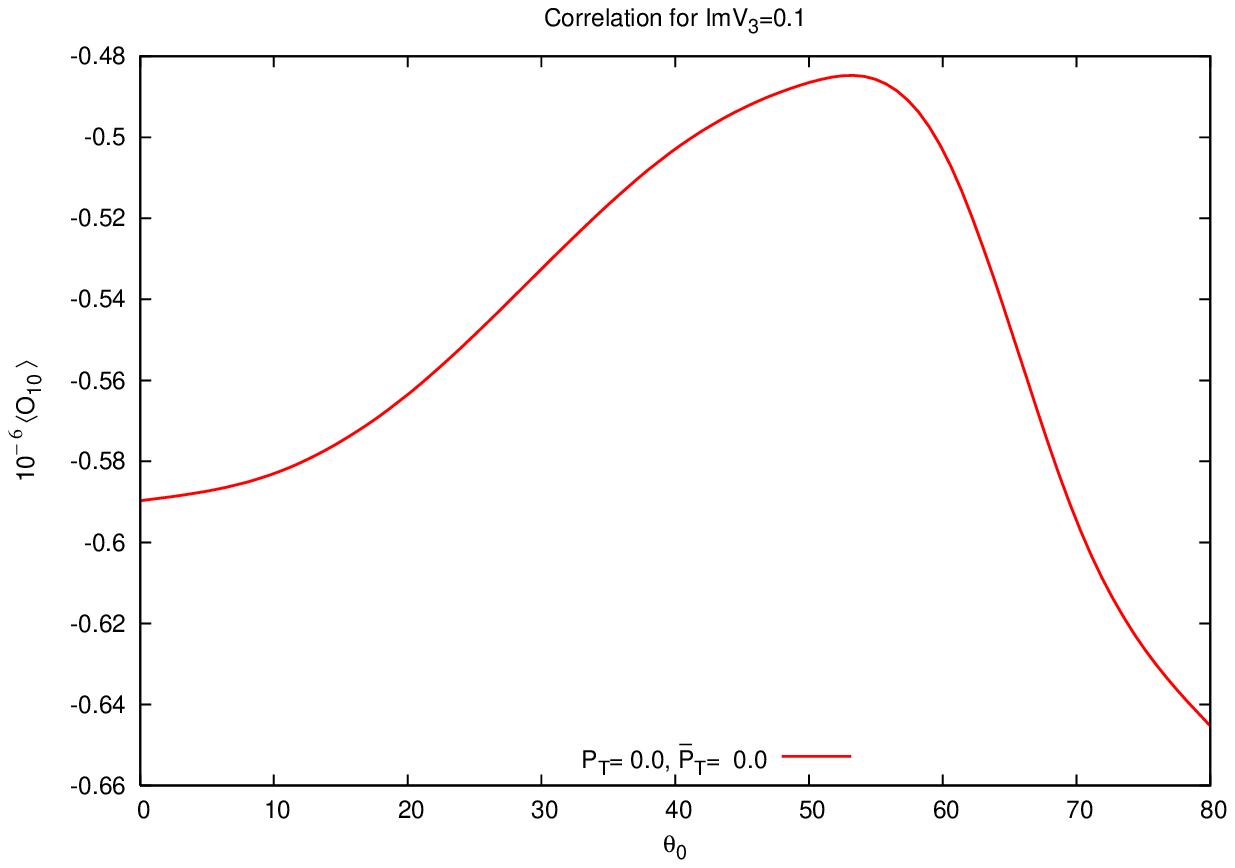}
\caption{The expectation value of $O_{10}$ (scaled down by an
appropriate factor) in GeV$^2$ with transversely 
polarized beams with Im~$V_3=0.1$. The remaining form factors are
zero.}\label{fig:10t10}
\end{figure}
\begin{figure}[!htb]
\centering
\epsfig{height=6.5cm,file=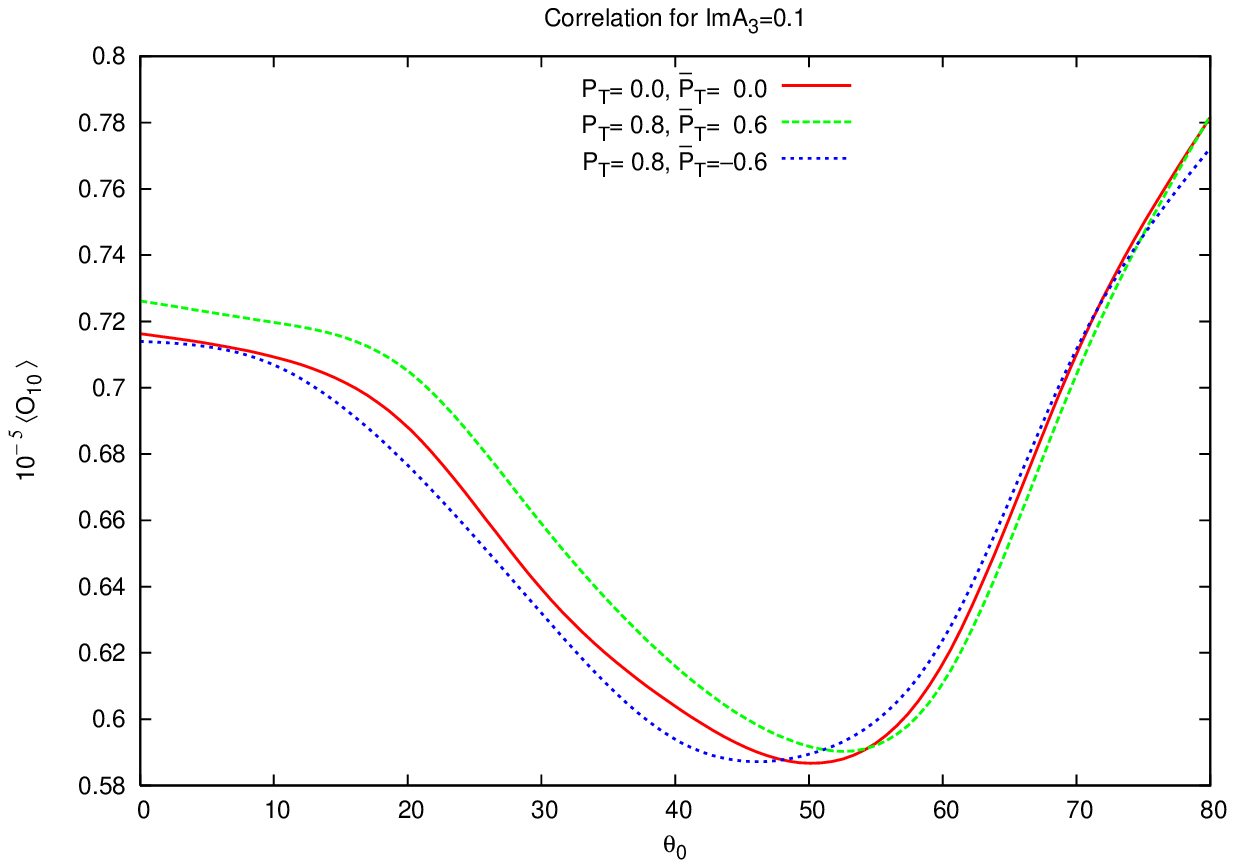}
\caption{The expectation value of $O_{10}$ (scaled down by an
appropriate factor) in GeV$^2$ with transversely 
polarized beams with Im~$A_3=0.1$. The remaining form factors are
zero.}\label{fig:10t12}
\end{figure}
Figs. \ref{fig:6t9}--\ref{fig:10t12} show the expectation values of
different observables for the case of transverse polarization, again,
choosing one form factor nonzero at a time. The
particulars of the nonzero form factors used are given in the respective
figure captions. 

The correlations are by and large weakly dependent on the cut-off. They
vary by 10 to 20\% over the whole range of $\theta_0$.

We do not show the plot of expectation values of $O_2$ because they turn
out to be independent of transverse polarization. Hence the expectation
values can be read off from Figs. \ref{fig:2l9} and \ref{fig:2l11}.
This does not, however, mean that it is not useful to use transverse
polarization in this case. The expectation value of the square of $O_2$,
which has a bearing on the sensitivity according to eq.
(\ref{eq:limits}), turns out to be dependent on the polarization. 
Thus, as will be seen below, the limit that can be
obtained on Re~$A_3$ from a measurement of $\langle O_2 \rangle$ can
improve with transverse polarization.

Similarly, the expectation values of $O_{10}$ are very weakly
dependent on transverse polarization.
In Fig. \ref{fig:10t10}, all the three curves
corresponding to the three polarization choices lie almost on top of one
another. So only the curve corresponding to $P_T=\overline P_T = 0$ is
shown.

%\documentclass[12pt]{article}
%\begin{document}
\begin{table}[htb]
\begin{tabular}{|c|c|c|c|c|c|}
\hline
&&Form&\multicolumn{3}{|c|}{Limits for beam polarizations}\\ \cline{4-6}
Symbol & Observable & Factor &
 $P_L=0$ & $P_L=+0.8$ & $P_L=+0.8$\\
&&& $\overline P_L=0$ & $\overline P_L=+0.6$ & $\overline P_L=-0.6$\\
\hline
$O_1$ & $(p_1 - p_2)\cdot q$   & Im~$V_3$ 
&$6.4\times 10^{-4}$ &
$2.6\times 10^{-4}$ & 
$7.2\times 10^{-5}$ \\
&& Im~$A_3$ &
$7.7\times 10^{-5}$ &
$7.3\times 10^{-5}$ & 
$6.6\times 10^{-5}$ \\
%%%%%%%%%%%%%%%%%%%%%%%%%%%%%%%%%%
\hline
$O_2$ &$(\vec p_{l^-} \times \vec p_{l^+})_z$ &Re~$V_3$ &
$2.2\times 10^{-3}$ &
$9.1\times 10^{-4}$ &
$2.5\times 10^{-4}$      \\
&& Re~$A_3$ & 
$2.6\times 10^{-4}$ &
$2.5\times 10^{-4}$ &
$2.3\times 10^{-4}$   \\
%%%%%%%%%%%%%%%%%%%%%%%%%%%%%%%%%%
\hline
$O_3$ & $(p_1 - p_2)\cdot q$
& Im~$V_2$ &
$5.2\times 10^{-3}$ &
$2.1\times 10^{-3}$ &
$5.9\times 10^{-4}$ \\
& $\times\, (\vec p_{l^-} \times \vec p_{l^+})_z$ 
& Im~$A_2$ & 
$6.2\times 10^{-4}$ &
$5.9\times 10^{-4}$ &
$5.5\times 10^{-4}$  \\
%%%%%%%%%%%%%%%%%%%%%%%%%%%%%%%%%%%
\hline
$O_5$ & $\! \! (p_1\! -\! p_2)\!\cdot \! ( p_{l^-}\! -\! p_{l^+})\!\!  $
& Re~$V_3$ &
$2.3\times 10^{-4}$ &
$2.3\times 10^{-4}$ &
$2.1\times 10^{-4}$ \\
& $ \times\, (\vec p_{l^-} \times \vec p_{l^+})_z$
& Re~$A_3$ &
$2.0\times 10^{-3}$ &
$8.2\times 10^{-4}$ &
$2.2\times 10^{-4}$  \\
%%%%%%%%%%%%%%%%%%%%%%%%%%%%%%%%%%%
\hline
$O_{10}$ & $P\cdot  ( p_{l^-} - p_{l^+}) $
& Im~$V_3$ &
$7.3\times 10^{-4}$ &
$6.9\times 10^{-4}$ &
$6.4\times 10^{-4}$ \\
& 
& Im~$A_3$ &
$6.0\times 10^{-3}$ &
$2.5\times 10^{-3}$ &
$6.8\times 10^{-4}$  \\
%%%%%%%%%%%%%%%%%%%%%%%%%%%%%%%%%%%
\hline

\end{tabular}
\caption{90\% C.L. limits on the form factors, chosen nonzero one at a time,
from the observables $O_1$, $O_2$ and $O_3$ with unpolarized and
longitudinally polarized beams.}\label{table:long}
\end{table}

\begin{table}[htb]
\begin{tabular}{|c|c|c|c|c|c|}
\hline
&&Form&\multicolumn{3}{|c|}{Limits for beam polarizations}\\ \cline{4-6}
Symbol & Observable &  Factor &
 $P_T=0$ & $P_T=+0.8$ & $P_T=+0.8$\\
&&& $\overline P_T=0$ & $\overline P_T=+0.6$ & $\overline P_T=-0.6$\\
\hline
$O_1$ & $(p_1 - p_2)\cdot q$   & Im~$V_3$ 
&$6.4\times 10^{-4}$ &
$8.9\times 10^{-4}$ & 
$1.7\times 10^{-4}$ \\
&& Im~$A_3$ &
$7.7\times 10^{-5}$ &
$1.1\times 10^{-4}$ & 
$2.0\times 10^{-5}$ \\
%%%%%%%%%%%%%%%%%%%%%%%%%%%%%%%%%%
\hline
$O_2$ &$(\vec p_{l^-} \times \vec p_{l^+})_z$ &Re~$V_3$ &
$2.2\times 10^{-3}$ &
$3.0\times 10^{-3}$ &
$5.7\times 10^{-4}$      \\
&& Re~$A_3$ & 
$2.6\times 10^{-4}$ &
$3.6\times 10^{-4}$ &
$6.9\times 10^{-5}$   \\
%%%%%%%%%%%%%%%%%%%%%%%%%%%%%%%%%%
\hline
$O_3$ & $(p_1 - p_2)\cdot q$
& Im~$V_2$ &
$5.2\times 10^{-3}$ &
$7.0\times 10^{-3}$ &
$1.3\times 10^{-3}$ \\
& $ \times \, (\vec p_{l^-} \times \vec p_{l^+})_z$ 
& Im~$A_2$ & 
$6.2\times 10^{-4}$ &
$8.6\times 10^{-4}$ &
$1.6\times 10^{-4}$  \\
%%%%%%%%%%%%%%%%%%%%%%%%%%%%%%%%%%%
\hline
$O_5$ & $\!\! (p_1\! -\! p_2)\!\cdot \! (p_{l^-}\! -\! p_{l^+})\!\! $
& Re~$V_3$ &
$2.3\times 10^{-4}$ &
$3.4\times 10^{-4}$ &
$6.2\times 10^{-5}$ \\
& $\times \,  (\vec p_{l^-} \times \vec p_{l^+})_z$
& Re~$A_3$ &
$2.0\times 10^{-3}$ &
$2.7\times 10^{-3}$ &
$5.2\times 10^{-4}$  \\
%%%%%%%%%%%%%%%%%%%%%%%%%%%%%%%%%%%
\hline
$O_6$ & $q_x\, q_y\, q_z$
& Re~$V_3$ &
&
$3.1\times 10^{-4}$ &
$5.7\times 10^{-5}$ \\
& 
& Re~$A_3$ &
&
$2.6\times 10^{-3}$ &
$4.8\times 10^{-4}$  \\
%%%%%%%%%%%%%%%%%%%%%%%%%%%%%%%%%%%
\hline
$O_7$ & $( q_x^2 -   q_y^2)\, q_z$
& Im~$V_3$ &
&
$2.6\times 10^{-3}$ &
$4.8\times 10^{-4}$ \\
& 
& Im~$A_3$ &
&
$3.1\times 10^{-4}$ &
$5.7\times 10^{-5}$  \\
%%%%%%%%%%%%%%%%%%%%%%%%%%%%%%%%%%%
\hline
$O_8$ & $(\vec p_{l^-} - \vec p_{l^+})_x$
& Re~$V_3$ &
&
$2.3\times 10^{-3}$ &
$4.3\times 10^{-4}$ \\
& 
$\times \, (\vec p_{l^-} - \vec p_{l^+})_y\, q_z$
& Re~$A_3$ &
&
$2.1\times 10^{-2}$ &
$3.5\times 10^{-3}$  \\
%%%%%%%%%%%%%%%%%%%%%%%%%%%%%%%%%%%
\hline
$O_9$ & $q_x\, q_y\, (\vec p_{l^-} - \vec p_{l^+})_z$
& Im~$V_2$ &
&
$1.4\times 10^{-2}$ &
$2.3\times 10^{-3}$ \\
&
& Im~$A_2$ &
&
$1.7\times 10^{-3}$ &
$3.1\times 10^{-4}$  \\
%%%%%%%%%%%%%%%%%%%%%%%%%%%%%%%%%%%
\hline
$O_{10}$ & $P\cdot  ( p_{l^-} - p_{l^+}) $
& Im~$V_3$ &
$7.3\times 10^{-4}$ &
$1.0\times 10^{-3}$ &
$1.8\times 10^{-4}$ \\
&
& Im~$A_3$ &
$6.0\times 10^{-3}$ &
$8.2\times 10^{-3}$ &
$1.6\times 10^{-3}$  \\
%%%%%%%%%%%%%%%%%%%%%%%%%%%%%%%%%%%
\hline
\end{tabular}
\caption{90\% C.L. limits on form factors, chosen nonzero one at a time,
from the observables $O_1$, $O_2$ and $O_3$ with unpolarized and
transversely polarized beams. The cut-off angle $\theta_0$ is chosen to
be $30^{\circ}$, except for the case of $O_8$, for which it is
$10^{\circ}$.}\label{table:trans}
\end{table}
%\end{document}

The 90\% CL limits that may be obtained at the ILC  running at
$\sqrt{s}=500$~GeV with an integrated luminosity of 500 fb$^{-1}$ have
been calculated for various cases using eq. (\ref{eq:limits}). The
results are presented in Table~\ref{table:long} for longitudinal
polarization and in Table~\ref{table:trans} for transverse polarization.

Table~\ref{table:long} shows that the best limits on all form factors
using unpolarized or longitudinally polarized beams are of the order of
about $10^{-4}$. Moreover, the best limits for any form factor 
are obtained for a suitable choice of observable when the
electron and positron polarizations are opposite in sign.

Transverse polarization allows more observables to be constructed
because of an additional azimuthal angle becomes available. Thus,
Table~\ref{table:trans} has more entries than Table~\ref{table:long}.
The entries which are blank 
in Table~\ref{table:trans} for limits in the unpolarized case are meant
to imply that the corresponding correlation is zero.
These correlations which are zero in the unpolarized and longitudinal
polarization cases are nonzero with transversely polarized beams. 
They thus give an independent measurement of certain form factors.
Again, the orders of  magnitude of the best limits are still around
10$^{-4}$, with some cases when it is possible to go down to about
$2\times 10^{-5}$.

We now make some detailed observations on our results.

$O_3 \equiv (p_1 - p_2)\cdot q (\vec p_{l^-} \times \vec p_{l^+})$ gives zero
expectation value  in the absence of polarization if only Im~$V_1$ and Im~$A_1$
are nonzero. This also means that this correlation is zero for SM. 
This continues to be true for longitudinal polarization because, as can been
seen explicitly from the expression for the cross section, the cross
section does not depend on Im~$V_1$ and Im~$A_1$. 
Hence it can be used to determine Im~$V_2$ and Im~$A_2$
with unpolarized beams. 

With transverse polarization, the cross section does 
depend on imaginary parts of $V_1$, and $A_1$. However, the
correlation vanishes when only Im~$V_1$ and Im~$A_1$ are nonzero.

$O_4 \equiv (p_1 - p_2)\cdot ( p_{l^-} - p_{l^+})$, on the other hand, 
has nonzero expectation value for Re~$V_1$ and Re~$A_1$, and hence for
SM. It is thus more difficult to use this correlation to study new
physics, though not impossible. We do not consider this correlation
here.

$O_6$, $O_7$, $O_8$ and $O_9$
have zero expectation values with unpolarized beams. 
However, good
sensitivity is obtained with transversely polarized beams. 
Of these, $O_6$ and $O_7$ have zero expectation values 
even with transverse polarization,
when the combination of couplings corresponds to a CP-violating $ZZH$
vertex. Thus observation of nonzero expectation value for $O_6$ would
signal unambiguously the presence of four-point $eeHZ$ couplings we
consider here.

The correlation $O_8$, which is proportional to $\langle E_{l^-} -
E_{l^+} \rangle$, shows strong dependence on the cut-off angle
$\theta_0$, changing sign around 25$^{\circ}$. Consequently, the
sensitivity is low for $\theta_0=30^{\circ}$. We have therefore chosen
to evaluate the limits at $\theta_0=10^{\circ}$ for this case.

We have also varied the Higgs mass up to 350 GeV and calculated all the above
correlations. However, the dependence on Higgs mass is rather weak. Thus
our results hold reasonably well even for larger Higgs
masses.

\section{Conclusions and discussion}

We have parametrized the amplitude 
for the process $e^+e^- \to HZ$ 
by means of form factors, 
using only Lorentz invariance,
treating separately
the chirality-conserving and chirality-violating cases. We then
calculated the differential cross section for the process $e^+e^- \to
HZ \to H l^+l^-$ in terms of these form factors for polarized beams 
for the chirality-conserving case.
The motivation was to determine the  extent to which 
longitudinal and transverse polarizations
can help in an independent determination of the various form factors.

In our earlier work \cite{plb}, where we considered only $Z$
angular distributions,
we found that in the presence of transverse polarization, there is a
CP-odd and T-odd contribution to the angular distribution. The coupling
combinations this term depends on cannot be determined using
longitudinally polarized beams. However, by looking at the 
distributions of charged leptons arising from the decay of the $Z$, we
can construct CP- and T-odd correlations even in the absence of
transverse polarization.

There do exist transverse-polarization
dependent correlations which do not arise when only $VVH$ type of couplings
are considered, as for example $O_6$ and $O_7$. 
These correlations, if observed, would be a unique signal
of CP-violating four-point interaction.

It should be emphasized that our results and conclusions are dependent
on
the assumption that the form factors are independent of $t$ and $u$. In
particular, the CP property of a given term in the distribution would
change if the corresponding form factor is an odd function of
$\cos\theta$. The reason is that $\cos\theta \equiv q\cdot
(p_2-p_1)/(\vert \vec q \vert s^{1/2})$ is odd under CP.

We have discussed limits on the couplings that would be expected from a
definite configuration of the linear collider. 
As for the 
CP-conserving couplings, limits may be obtained even from the existing
LEP data, which has excluded SM Higgs up to mass of about 114 GeV.
It should be borne in mind that the limits on these depend on the choice 
of $M$, the arbitrary parameter of dimension of mass that we introduced. 

We have looked at expectation values of observables taking only one form
factor nonvanishing at a time. It must be emphasized that it feasible to
extract information on form factors even when they are allowed to be
nonzero independently of one another by one of two methods: Either one
determines experimentally more than one correlation and solves the
linear simultaneous equations for the form factors or one combines
experimental information obtained with  unpolarized and polarized beams
to solve for the form factors. For details of this straightforward
procedure, see  \cite{poulose,sdr}.

As compared to the earlier study \cite{plb} in which only the $Z$ polar
and azimuthal angles were considered as measurable, the use of angular
variables of both leptons already presents certain new observables which
give new information with even unpolarized beams, though longitudinal
polarization can improve the sensitivity. The use of transverse
polarization in such a situation does not present as many advantages as
it did when only $Z$ variables were used in asymmetries. However, if use
is made of observables which involve the momentum of only one lepton at
a time, transverse polarizations would again prove advantageous.

We can compare the sensitivities obtained on including $Z$ decay with the
sensitivities obtained using angular distributions of the $Z$ itself,
which were studied in \cite{plb}. There, we had used $M=1000$~GeV,
whereas here we use $M=100$~GeV. Keeping in mind this difference, we find 
that it is possible to reach similar limits using suitable correlations
even in the present case.
Note that in \cite{plb} what was used was asymmetries, rather
than the expectation values used here.

We have also studied how our results are affected by more realistic
cuts, viz., cuts on the energy of the leptons and on their transverse
momentum. Since we already assume a cut on the lepton polar angles, it
is not necessary to study the effect of the cut on transverse momentum
separately from a cut on the energy.
We find that for cuts on the energies of both leptons of 20 (30) GeV
all relevant quantities (cross section, correlation and the limits on the
form factors change by about 5\%  (10\%). Thus our conclusions would
remain valid fairly accurately even with experimental cuts.  

We assumed that the Higgs can be detected with full efficiency. In
practice, of course, the detection of the Higgs would require putting
cuts on the Higgs decay products, leading to an efficiency factor less
than 1. This would make  our limits worse. On the other hand, some of
our correlations which involve only the sum of the momenta of the
leptons can also be extended to the case of hadronic decays of the $Z$.
In that case, the event sample would be larger, improving the limits.

One should keep in mind the possibility that electroweak radiative corrections,
which can be particularly large for transverse polarization
\cite{comelli}, can 
lead to quantitative changes in the above results.

Nevertheless, we would like to emphasize that our work contains 
full analytical expressions for charged-lepton distributions, and 
can prove a useful input to a more exhaustive
work taking into account, on the one hand, a variety of specific
models, and on the other hand more precise experimental constraints and
other practical considerations.

Though we have used SM couplings for the leading contribution of Fig.
\ref{fig:vvhptgraph}, as mentioned earlier, the analysis needs only
trivial
modification when applied to a model like MSSM or a multi-Higgs-doublet
model, and will be useful in such extensions of SM.
It is likely that such models will give rise to four-point contributions
through box diagrams or loop diagrams with a $t$-channel exchange of
particles. However, to our knowledge, such calculations are not
available for
CP-violating models. The interesting effects we have discussed would
make it useful to carry out such calculations.

\noindent Acknowledgment: This work was partly supported by the IFCPAR
project no. 3004-2.

%\cite{Kilian:1996ja}
%\bibitem{Kilian:1996ja}
W.~Kilian, M.~Kramer and P.~M.~Zerwas,
%``Anomalous couplings in the Higgs-strahlung process,''
arXiv:hep-ph/9605437;
%%CITATION = HEP-PH 9605437;%%
  %``Anomalous couplings in the Higgs-strahlung process,''
  Phys.\ Lett.\  B {\bf 381}, 243 (1996)
  [arXiv:hep-ph/9603409].
  %%CITATION = PHLTA,B381,243;%%

\bibitem{gounaris}
%\cite{Gounaris:1995mx}
%\bibitem{Gounaris:1995mx}
G.~J.~Gounaris, F.~M.~Renard and N.~D.~Vlachos,
%``Tests of Anomalous Higgs Boson Couplings through $e~-e~+ \to HZ$
%and
%$H\gamma$,''
Nucl.\ Phys.\ B {\bf 459}, 51 (1996)
[arXiv:hep-ph/9509316].

\bibitem{hagiwara}
%\cite{Hagiwara:1993sw}
%\bibitem{Hagiwara:1993sw}
K.~Hagiwara and M.~L.~Stong,
%``Probing the scalar sector in e+ e- $\to$ f anti-f H,''
Z.\ Phys.\ C {\bf 62}, 99 (1994)
[arXiv:hep-ph/9309248].

\bibitem{kniehl1}
  V.~D.~Barger, K.~m.~Cheung, A.~Djouadi, B.~A.~Kniehl and P.~M.~Zerwas,
  %``Higgs bosons: Intermediate mass range at e+ e- colliders,''
  Phys.\ Rev.\  D {\bf 49}, 79 (1994)
  [arXiv:hep-ph/9306270].
  %%CITATION = PHRVA,D49,79;%%

\bibitem{kniehl2}
  K.~Hagiwara, S.~Ishihara, J.~Kamoshita and B.~A.~Kniehl,
  %``Prospects of measuring general Higgs couplings at e+ e- linear
  %colliders,''
  Eur.\ Phys.\ J.\  C {\bf 14}, 457 (2000)
  [arXiv:hep-ph/0002043].
  %%CITATION = EPHJA,C14,457;%%
 \bibitem{Skjold}
   A.~Skjold and P.~Osland,
%  %``Testing CP in the Bjorken process,''
   Nucl.\ Phys.\  B {\bf 453}, 3 (1995)
   [arXiv:hep-ph/9502283].
  %%CITATION = NUPHA,B453,3;%%

\bibitem{kile}
  J.~Kile and M.~J.~Ramsey-Musolf,
  %``Fermionic effective operators and Higgs production at a linear collider,''
  arXiv:0705.0554 [hep-ph].
  %%CITATION = ARXIV:0705.0554;%%

\bibitem{basdr}
B.~Ananthanarayan and S.~D.~Rindani,
%Transverse beam polarization and CP violation in
%$e^+ e^- \to \gamma Z$ with
%  contact interactions,}
Phys.\ Lett.\ B {\bf 606}, 107 (2005) 
[arXiv:hep-ph/0410084];
%%CITATION = HEP-PH 0410084
%``New physics in e+ e- $\to$ Z gamma with polarized beams,''
JHEP {\bf 0510}, 077 (2005)
[arXiv:hep-ph/0507037].
%%CITATION = HEP-PH 0507037;%%

\bibitem{AL1}
%\cite{Abraham:1998cm}
K.~J.~Abraham and B.~Lampe,
%``Possible nonstandard effects in Z gamma events at LEP2,''
Phys.\ Lett.\ B {\bf 446}, 163 (1999).
[arXiv:hep-ph/9810205].
%%CITATION = HEP-PH 9810205;%%

\bibitem{AL2}
K.~J.~Abraham and B.~Lampe,
%Description of possible CP effects in $b\bar{b} \gamma$ events at
%LEP,}
Phys.\ Lett.\ B {\bf 326}, 175 (1994).
%%CITATION = PHLTA,B326,175;%%
%%CITATION = HEP-PH 9509316;%%
%%CITATION = HEP-PH 9309248;%%

\bibitem{lepto}
%\cite{Rindani:2004ue}
%\bibitem{Rindani:2004ue}
S.~D.~Rindani,
%``Transverse beam polarization and limits on leptoquark couplings
%in e+  e-
%$\to$ t anti-t,''
Phys.\ Lett.\ B {\bf 602}, 97 (2004)
[arXiv:hep-ph/0408083].
%%CITATION = HEP-PH 0408083;%%

\bibitem{plb}
  K.~Rao and S.~D.~Rindani,
  %``Probing CP-violating contact interactions in e+ e- --> H Z with
  %polarized
  %beams,''
  Phys.\ Lett.\  B {\bf 642}, 85 (2006)
  [arXiv:hep-ph/0605298].
  %%CITATION = PHLTA,B642,85;%%

\bibitem{chen}
  C.~M.~J.~Chen, J.~W.~J.~Chen and W.~Y.~P.~Hwang,
  %``Higgs boson production via the Bjorken process e+ e- $\to$ H0 mu+ mu- at
  %high-energy e+ e- colliders,''
  Phys.\ Rev.\  D {\bf 50}, 4485 (1994).
  %%CITATION = PHRVA,D50,4485;%%

%\cite{Skjold:1995jp}
%\bibitem{Skjold:1995jp}
%  A.~Skjold and P.~Osland,
%  %``Testing CP in the Bjorken process,''
%  Nucl.\ Phys.\  B {\bf 453}, 3 (1995)
%  [arXiv:hep-ph/9502283].
  %%CITATION = NUPHA,B453,3;%%

\bibitem{mahlon}
  G.~Mahlon and S.~J.~Parke,
  %``Using spin correlations to distinguish Z h from Z A at the international
  %linear collider,''
  Phys.\ Rev.\  D {\bf 74}, 073001 (2006)
  [arXiv:hep-ph/0606052].
  %%CITATION = PHRVA,D74,073001;%%

\bibitem{gudi}
G.~Moortgat-Pick {\it et al.},
%``The role of polarized positrons and electrons in revealing
%fundamental
%interactions at the linear collider,''
arXiv:hep-ph/0507011.
%%CITATION = HEP-PH 0507011;%%

\bibitem{rizzo}
T.~G.~Rizzo,
%``Transverse polarization signatures of extra dimensions at linear
%colliders,''
JHEP {\bf 0302}, 008 (2003)
[arXiv:hep-ph/0211374];
%%CITATION = HEP-PH 0211374;%%
%``More transverse polarization signatures of extra dimensions at
%linear
%colliders,''

J.~Fleischer, K.~Kolodziej and F.~Jegerlehner,
%``Transverse versus longitudinal polarization effects in e+ e- $\to$
%W+ W-,''  
Phys.\ Rev.\ D {\bf 49}, 2174 (1994);
%%CITATION = PHRVA,D49,2174;%%

M.~Diehl, O.~Nachtmann and F.~Nagel,
%``Probing triple gauge couplings with transverse beam polarisation in
%e+  e-
%$\to$ W+ W-,''
Eur.\ Phys.\ J.\ C {\bf 32}, 17 (2003)
[arXiv:hep-ph/0306247];
%%CITATION = HEP-PH 0306247;%%

S.~Y.~Choi, J.~Kalinowski, G.~Moortgat-Pick and P.~M.~Zerwas,
%``Analysis of the neutralino system in supersymmetric theories,''
Eur.\ Phys.\ J.\ C {\bf 22}, 563 (2001)
[Addendum-ibid.\ C {\bf 23}, 769 (2002)]
[arXiv:hep-ph/0108117].
  %%CITATION = HEP-PH 0108117;%%

\bibitem{basdrtt}
B.~Ananthanarayan and S.~D.~Rindani,
%``CP violation at a linear collider with transverse polarization,''
Phys.\ Rev.\ D {\bf 70}, 036005 (2004)
[arXiv:hep-ph/0309260];
%%CITATION = HEP-PH 0309260;%%

\bibitem{basdrzzg}
B.~Ananthanarayan, S.~D.~Rindani, R.~K.~Singh and A.~Bartl,
%``Transverse beam polarization and CP-violating triple gauge boson
%couplings  %in e+ e- $\to$ gamma Z,''
Phys.\ Lett.\ B {\bf 593}, 95 (2004)
[Erratum-ibid.\ B {\bf 608}, 274 (2005)]
[arXiv:hep-ph/0404106];
%%CITATION = HEP-PH 0404106;%%

J.~Kalinowski,
%``Testing CP violation and universal extra dimensions at future
%colliders,''
arXiv:hep-ph/0410137;
%%CITATION = HEP-PH 0410137;%%

P.~Osland and N.~Paver,
%``Positron polarization at the International Linear Collider,''
arXiv:hep-ph/0507185.
%%CITATION = HEP-PH 0507185;%%

\bibitem{bartl}
A.~Bartl, K.~Hohenwarter-Sodek, T.~Kernreiter and H.~Rud,
%``CP sensitive observables in chargino production with transverse
%e+-  beam
%polarization,''
Eur.\ Phys.\ J.\ C {\bf 36}, 515 (2004)
[arXiv:hep-ph/0403265];
%%CITATION = HEP-PH 0403265;%%

A.~Bartl, H.~Fraas, S.~Hesselbach, K.~Hohenwarter-Sodek, T.~Kernreiter
and G.~Moortgat-Pick,
%``CP-odd observables in neutralino production with transverse e+ and
%e- beam
%polarization,''
JHEP {\bf 0601}, 170 (2006)
[arXiv:hep-ph/0510029];
%%CITATION = HEP-PH 0510029;%%

S.~Y.~Choi, M.~Drees and J.~Song,
%``Neutralino production and decay at an e+ e- linear collider with
%transversely polarized beams,''
arXiv:hep-ph/0602131;
%%CITATION = HEP-PH 0602131;%%

A.~Bartl, K.~Hohenwarter-Sodek, T.~Kernreiter and O.~Kittel,
%``CP asymmetries with Longitudinal and Transverse Beam Polarizations in
%Neutralino Production and Decay into the Z^0 Boson at the ILC,''
arXiv:0706.3822 [hep-ph];
%%CITATION = ARXIV:0706.3822;%%

A.~Bartl, H.~Fraas, K.~Hohenwarter-Sodek, T.~Kernreiter,
G.~Moortgat-Pick and A.~Wagner,
%``Selectron production at an e- e- linear collider with transversely
%polarized beams,''
Phys.\ Lett.\  B {\bf 644}, 165 (2007)
[arXiv:hep-ph/0610431].
%%CITATION = PHLTA,B644,165;%%

\bibitem{Rindani:2004wr}
S.~D.~Rindani,
%``Role of transverse polarization in constraining new physics,''
arXiv:hep-ph/0409014.
%%CITATION = HEP-PH 0409014;%%

\bibitem{form} 
  J.~A.~M.~Vermaseren,
  %``New features of FORM,''
  arXiv:math-ph/0010025.
  %%CITATION = MATH-PH/0010025;%%

\bibitem{sdrCP} See, for example, S.D.~Rindani, 
  %``CP symmetry and its violation in fundamental interactions,''
  Pramana {\bf 49}, 81 (1997);
  %%CITATION = PRAMC,49,81;%%
  %``CP violation at colliders,''
  Pramana {\bf 45}, S263 (1995)
  [arXiv:hep-ph/9411398];
  %%CITATION = PRAMC,45,S263;%%

  G.~C.~Branco, L.~Lavoura and J.~P.~Silva,
``CP violation,''
%\href{http://www.slac.stanford.edu/spires/find/hep/www?irn=4269535}{SPIRES entry}
{\it  Oxford, UK: Clarendon (1999) 511 p}

\bibitem{hikasa} K.-i.~Hikasa,
%``Transverse Polarization Effects In E+ E- Collisions: The Role Of
%Chiral
%Symmetry,''
Phys.\ Rev.\ D {\bf 33}, 3203 (1986).
%%CITATION = PHRVA,D33,3203;%%

\bibitem{poulose} 
  P.~Poulose and S.~D.~Rindani,
  %``Simple decay-lepton asymmetries in polarized e+ e- --> t anti-t and
  %CP-violating dipole couplings of the top quark,''
  Phys.\ Lett.\  B {\bf 383}, 212 (1996)
  [arXiv:hep-ph/9606356];
  %%CITATION = PHLTA,B383,212;%%
  %``Decay-lepton angular distribution in polarized e+ e- --> t anti-t
%and
  %CP-violating dipole couplings of the top quark,''
  Phys.\ Rev.\  D {\bf 54}, 4326 (1996)
  [Erratum-ibid.\  D {\bf 61}, 119901 (2000)]
  [arXiv:hep-ph/9509299];
  %%CITATION = PHRVA,D54,4326;%%
  %``CP violating asymmetries in e+ e- $\to$ t anti-t with
  %longitudinally
  %polarized electrons,''
  Phys.\ Lett.\  B {\bf 349}, 379 (1995)
  [arXiv:hep-ph/9410357];
  %%CITATION = PHLTA,B349,379;%%

 F.~Cuypers and S.~D.~Rindani,
  %``Top quarks and CP violation in polarized e+ e- collisions,''
  Phys.\ Lett.\  B {\bf 343}, 333 (1995)
  [arXiv:hep-ph/9409243].
  %%CITATION = PHLTA,B343,333;%%

\bibitem{sdr}
  S.~D.~Rindani,
  %``Single decay-lepton angular distributions in polarized e+ e- --> t
%anti-t
  %and simple angular asymmetries as a measure of CP-violating top
%dipole
  %couplings,''
  Pramana {\bf 61}, 33 (2003)
  [arXiv:hep-ph/0304046].
  %%CITATION = PRAMC,61,33;%%

\bibitem{comelli}
P.~Ciafaloni, D.~Comelli and A.~Vergine,
%``Sudakov electroweak effects in transversely polarized beams,''
JHEP {\bf 0407} (2004) 039
[arXiv:hep-ph/0311260].
%%CITATION = HEP-PH 0311260;%%

\end{document}